\documentclass[10pt]{article}
\usepackage{graphicx}
\usepackage{amsmath}
\usepackage{amssymb}
\usepackage{caption2}
\usepackage{booktabs}
\usepackage{setspace}
\begin{document}
\bibliographystyle{prsty}

\begin{center}
{\large {\bf \sc{Two-body strong decays  of the pseudoscalar hidden-charm tetraquark states via the QCD sum rules}}} \\[2mm]

Yu-Hang Xu,  Zhi-Gang Wang\footnote{E-mail: zgwang@aliyun.com.  } \\
Department of Physics, North China Electric Power University, Baoding 071003, P. R. China
\end{center}

\begin{abstract}
In this work, we study the properties of the pseudoscalar hidden-charm tetraquark states by analyzing  their two-body strong decays via the QCD sum rules based on rigorous quark-hadron duality. We take into account the vacuum condensates up to dimension 5 on the QCD side, and obtain the hadronic coupling constants. At last, we obtain the total decay widths   $\Gamma(Z_{c}^{-}) = 326.20^{+4.26}_{-3.11}$ MeV  and $\Gamma(Z_{c}^{+}) = 91.84^{+0.96}_{-0.76}$ MeV, respectively, where the $Z_{c}^{+}$($J^{PC}=0^{-+}$) and $Z_{c}^{-}$($J^{PC}=0^{--}$) denote the pseudoscalar hidden-charm tetraquarks with the diquark-antidiquark structures $[uc]_{A}[\bar{d}\bar{c}]_{V}-[uc]_{V}[\bar{d}\bar{c}]_{A}$  and $[uc]_{A}[\bar{d}\bar{c}]_{V}+[uc]_{V}[\bar{d}\bar{c}]_{A}$, respectively.
\end{abstract}

\section{Introduction}
Since observation of the $X(3872)$  by the Belle collaboration in 2003  \cite{Belle:2003nnu},  the Belle, BESIII  and LHCb collaborations  have observed and confirmed   many $X$, $Y$ and $Z$ states, such as the $Z_c(3900)$, $Y(4500)$, $Y(4660)$, etc, the exotic hadrons have attracted considerable attentions \cite{Review-penta-mole-ZhSL-RPT,Review-penta-Esposito-RPT,Review-penta-Ali-PPNP,
Review-penta-mole-GuoFK-RMP,Review-penta-mole-LiuYR-PPNP,
Review-penta-mole-Brambilla-RPT,Review-mole-GuoFK-CTP,
Review-mole-WangB-PRT,Review-mole-GengLS-PRT,WangZG-review}. Those  exotic hadrons cannot  be well understood in the traditional quark model, the theoretical researchers have proposed several models to explain their nature, such as the molecular states \cite{Guo:2008zg,HeJun-Zc3900-mole,WangZG-mole-formula,LiuYL-mole-JPG2014,
Albuquerque:2011ix,Wang:2009hi}, multiquark states \cite{Ebert:2008kb,Chen:2010ze,Wang:2017lot,Wang:2019iaa,Yang:2025lef,Zhang:2010mw,Sundu:2018toi,Di:2018yoc,Wang:2018rfw,Wang:2013exa,Wang:2021qus,Wang:2023jaw}, hadrocharmonia \cite{Brambilla:2017ffe}, hybrids \cite{Varma:2018mmi,Yoo:2022erv,Yoo:2023spi}, etc.

The $X$, $Y$ and $Z$ states with exotic quantum numbers, such as $J^{PC}=0^{--}$, $1^{-+}$, etc,   cannot be the conventional quark-antiquark mesons and  are of extremely great interest. Many  theoretical works  have predicted
the $J^{P}=0^{-}$ states, including compact tetraquarks \cite{Wang:2021lkg,Chen:2010jd}, hadronic molecules \cite{Dong:2024fjk}, hybrids \cite{Huang:2016rro,Liu:2005rc} and glueballs \cite{Pimikov:2016pag,Qiao:2014vva,WangZG-PRD-2025-hybrid}, however, no experimental candidate  has been observed. We should carry out systematic theoretical studies in order to guide  experimental researches in the future.

The strong decays of the hidden-charm tetraquark states are expected to take place through the Okubo-Zweig-Iizuka super-allowed fall-apart mechanism without annihilating and creating a quark-antiquark pair. In Ref.\cite{Wang:2021lkg}, we constructed the diquark-antidiquark type   currents  without introducing explicit P-waves, and calculated the mass spectrum of the pseudoscalar hidden-charm tetraquarks via the QCD sum rules in our unique scheme, and obtained the lowest mass about $4.56 \pm 0.08~\mathrm{GeV}$ for the $c\bar{c}u\bar{d}$ state.

However, the mass alone only leads to a crude assessment. In this work, we study  the hadronic coupling constants in the two-body strong decays of the  hidden-charm tetraquark states with the $J^{PC}=0^{-+}$ and $0^{--}$ via the QCD sum rules based on the rigorous quark-hadron duality, which allows us to predict the partial and total decay widths.   At the first step, we study the $[uc]_{A}[\bar{d}\bar{c}]_{V}-[uc]_{V}[\bar{d}\bar{c}]_{A}$  and $[uc]_{A}[\bar{d}\bar{c}]_{V}+[uc]_{V}[\bar{d}\bar{c}]_{A}$  tetraquark states and  denote them as $Z_{c}^{{+}}$ and $Z_{c}^{{-}}$, respectively.  In calculations,  we take account of both the connected and disconnected Feynman diagrams to ensure accuracy.

The article is organized as follows: in Section 2, we obtain the QCD sum rules for the hadronic coupling constants; in Section 3, we present the numerical results and  discussions; finally,  we summarize our observations.

\section{ QCD sum rules for the hadronic coupling constants }
Firstly, let us write down the three-point correlation functions,

\begin{eqnarray}\label{CF-1}
\Pi^{\chi_{c1}\rho }_{-,\alpha\beta}(p,q)& =& i^2 \int d^4 x d^4 y e^{ip\cdot x} e^{iq\cdot y} \langle0| T\left\{J^{\chi_{c1}}_{\alpha}(x) J^{\rho}_{\beta}(y) 	 J_{-}^{\dagger}(0)\right\}|0\rangle \, , \nonumber \\[4pt]
\Pi^{\eta_{c}\rho }_{-,\alpha}(p,q) &=& i^2 \int d^4 x d^4 y e^{ip\cdot x} e^{iq\cdot y} \langle0| T\left\{J^{\eta_c}(x) J^{\rho}_{\alpha}(y) J_{-}^{\dagger}(0)\right\}|0\rangle  \, , \nonumber \\[4pt]
\Pi^{J/\psi a_1 }_{-,\alpha\beta}(p,q)& =& i^2 \int d^4 x d^4 y e^{ip\cdot x} e^{iq\cdot y} \langle0| T\left\{J^{J/\psi}_{\alpha}(x) J^{a_1}_{\beta}(y) J_{-}^{\dagger}(0)\right\}|0\rangle \, ,\nonumber \\[4pt]
\Pi^{J/\psi\pi }_{-,\alpha}(p,q) &=& i^2 \int d^4 x d^4 y e^{ip\cdot x} e^{iq\cdot y} \langle0| T\left\{J^{J/\psi}_{\alpha}(x) J^{\pi}(y) J_{-}^{\dagger}(0)\right\}|0\rangle \, ,\nonumber \\[4pt]
\Pi^{D\bar{D}_0 }_{-}(p,q) &=& i^2 \int d^4 x d^4 y e^{ip\cdot x} e^{iq\cdot y} \langle0|T\left\{J^{D}(x) J^{\bar{D}_0}(y) J_{-}^{\dagger}(0)\right\}|0\rangle \,\nonumber \\[4pt]
\Pi^{D^*\bar{D}_1 }_{-,\alpha\beta}(p,q)& =& i^2 \int d^4 x d^4 y e^{ip\cdot x} e^{iq\cdot y} \langle0|T\left\{J^{D^*}_{\alpha}(x) J^{\bar{D}_1}_{\beta}(y) J_{-}^{\dagger}(0)\right\}|0\rangle \, , \nonumber \\[4pt]
\Pi^{D^*\bar{D} }_{-,\alpha}(p,q) &=& i^2 \int d^4 x d^4 y e^{ip\cdot x} e^{iq\cdot y} \langle0| T\left\{J^{D^*}_{\alpha}(x) J^{\bar{D}}(y) J_{-}^{\dagger}(0) \right\}|0\rangle \, ,
\end{eqnarray}

\begin{eqnarray}\label{CF-2}
\Pi^{\chi_{c0}\pi }_{+}(p,q) &=& i^2 \int d^4 x d^4 y e^{ip\cdot x} e^{iq\cdot y} \langle0| T\left\{J^{\chi_{c0}}(x) J^{\pi}(y) J_{+}^{\dagger}(0) \right\}|0\rangle \, ,\nonumber \\[4pt]
\Pi^{\eta_c a_0 }_{+}(p,q) &=& i^2 \int d^4 x d^4 y e^{ip\cdot x} e^{iq\cdot y} \langle0| T\left\{J^{\eta_c}(x) J^{a_0}(y) J_{+}^{\dagger}(0)\right\}|0\rangle \, ,\nonumber \\[4pt]
\Pi^{J/\psi\rho }_{+,\alpha\beta}(p,q) &=& i^2 \int d^4 x d^4 y e^{ip\cdot x} e^{iq\cdot y} \langle0| T\left\{J^{J/\psi}_{\alpha}(x) J^{\rho}_{\beta}(y) J_{+}^{\dagger}(0) \right\}|0\rangle \, ,
\nonumber \\[4pt]
\Pi^{D\bar{D}_0 }_{+}(p,q) &=& i^2 \int d^4 x d^4 y e^{ip\cdot x} e^{iq\cdot y} \langle0|T\left\{J^{D}(x) J^{\bar{D}_0}(y) J_{+}^{\dagger}(0)\right\}|0\rangle \, ,\nonumber \\[4pt]
\Pi^{D^*\bar{D} }_{+,\alpha}(p,q) &=& i^2 \int d^4 x d^4 y e^{ip\cdot x} e^{iq\cdot y} \langle0|T\left\{J^{D^*}_{\alpha}(x) J^{\bar{D}_1}_{\beta}(y) J_{+}^{\dagger}(0)\right\}|0\rangle \, ,\nonumber \\[4pt]
\Pi^{D^*\bar{D} }_{+,\alpha}(p,q) &=& i^2 \int d^4 x d^4 y e^{ip\cdot x} e^{iq\cdot y} \langle0| T\left\{J^{D^*}_{\alpha}(x) J^{\bar{D}}(y) J_{+}^{\dagger}(0) \right\}|0\rangle \,\nonumber \\[4pt]
	\Pi^{D^*\bar{D}^* }_{+,\alpha\beta}(p,q) &=& i^2 \int d^4 x d^4 y e^{ip\cdot x} e^{iq\cdot y} \langle0| T\left\{J^{D^*}_{\alpha}(x) J^{\bar{D}^*}_{\beta}(y) J_{+}^{\dagger}(0) \right\}|0\rangle \, ,
\end{eqnarray}
where the currents
\begin{eqnarray}
J^{\chi_{c1}}_{\alpha}(x) &=& \bar{c}(x)\gamma_\alpha \gamma_5 c(x)\, , \nonumber\\[4pt]
J^{\rho}_{\alpha}(x)&=&\bar{d}(x)\gamma_\alpha  u(x)\, ,\nonumber\\[4pt]
J^{\eta_c}(x) &=& \bar{c}(x)i\gamma_5 c(x)\, , \nonumber\\[4pt]
J^{J/\psi}_{\alpha}(x) &=& \bar{c}(x)\gamma_\alpha c(x)\, , \nonumber\\[4pt]
J^{a_1}_{\alpha}(x) &=& \bar{d}(x)\gamma_\alpha \gamma_5 u(x)\, ,\nonumber\\[4pt]
J^{\pi}(x)&=&\bar{d}(x)i\gamma_5  u(x)\, ,\nonumber\\[4pt]
J^{\chi_{c0}}(x)&=&\bar{c}(x) c(x) \, ,\nonumber \\[4pt]
J^{a_0}(x)&=&\bar{d}(x)  u(x) \, ,
\end{eqnarray}
\begin{eqnarray}
J^{D}(x)&=&\bar{d}(x)i\gamma_{5} c(x) \, ,\nonumber \\[4pt]
J^{\bar{D}_0}(x)&=&\bar{c}(x) u(x) \, ,\nonumber \\[4pt]
J_{\alpha}^{D^*}(x)&=&\bar{d}(x)\gamma_{\alpha} c(x) \, ,\nonumber \\[4pt]
J_{\alpha}^{\bar{D}_1}(x)&=&\bar{c}(x)\gamma_\alpha \gamma_5 u(x) \, ,\nonumber \\[5pt]
J^{\bar{D}}(x)&=&\bar{c}(x)i\gamma_{5} u(x)  \, ,\nonumber \\[4pt]
J_{\beta}^{\bar{D}^*}(x)&=&\bar{c}(x)\gamma_{\beta} u(x) \, ,
\end{eqnarray}
the superscripts $\chi_{c1}$, $\rho$, $\eta_c$, $J/\psi$, $a_1$, $\pi$, $\chi_{c0}$, $a_0$, $D$, $\bar{D}_0$, $D^*$, $\bar{D}_1$,  $\bar{D}$, $\bar{D}^*$ denote the corresponding mesons, while the currents
\begin{eqnarray}	 J_{\pm}(x)&=&\frac{\varepsilon^{ijk}\varepsilon^{imn}}{\sqrt{2}}
\Big[u^{T}_j(x)C\gamma_{\mu}c_k(x) \bar{d}_{m}(x)\gamma_5\gamma^\mu C \bar{c}^{T}_n(x)\mp u^{T}_j(x)C\gamma_\mu\gamma_5 c_k(x)\bar{d}_{m}(x)\gamma^{\mu}C \bar{c}^{T}_n(x) \Big] \, ,\nonumber\\[5pt]
\end{eqnarray}
interpolate the hidden-charm tetraquark states with the  $J^{PC}=0^{-+}$ and $0^{--}$, respectively \cite{Wang:2021lkg}, the
subscripts $\pm$ stand for the positive and negative charge-conjugations, respectively.

On the phenomenological side, we insert a complete set of intermediate hadronic states with same quantum numbers as the currents into the three-point correlation functions in Eqs.\eqref{CF-1}-\eqref{CF-2}, then we explicitly isolate the contributions of the ground states \cite{SVZ79-1,SVZ79-2,Reinders:1984sr},	
\begin{eqnarray}
	\Pi^{\chi_{c1}\rho }_{-,\alpha\beta}(p,q) &=&\Pi_{\chi_{c1}\rho Z_c^{-}}(p^{\prime2},p^2,q^2)
	\,i g_{\alpha\beta}+\cdots\, , \nonumber\\[4pt]
	\Pi^{\eta_{c}\rho }_{-,\alpha}(p,q) &=&\Pi_{\eta_{c}\rho Z_c^{-}}(p^{\prime2},p^2,q^2)
	\,i p_\alpha+\cdots\, ,\nonumber\\[4pt]
	\Pi^{J/\psi a_1 }_{-,\alpha\beta}(p,q) &=&\Pi_{J/\psi a_1 Z_c^{-}}(p^{\prime2},p^2,q^2)
\,i g_{\alpha\beta}+\cdots\, ,\nonumber\\[4pt]
	\Pi^{J/\psi\pi }_{-,\alpha\beta}(p,q) &=&\Pi_{J/\psi\pi Z_c^{-}}(p^{\prime2},p^2,q^2)
\,i q_\alpha+\cdots\, ,\nonumber\\[4pt]
	\Pi^{D\bar{D}_0 }_{-}(p,q) &=&\Pi_{D\bar{D}_0 Z_c^{-}}(p^{\prime2},p^2,q^2)
	+\cdots\, ,\nonumber\\[4pt]
	\Pi^{D^*\bar{D}_1 }_{-,\alpha\beta}(p,q) &=&\Pi_{D^*\bar{D}_1 Z_c^{-}}(p^{\prime2},p^2,q^2)
\,i g_{\alpha\beta}+\cdots\, ,\nonumber\\[4pt]
	\Pi^{D^*\bar{D} }_{-,\alpha}(p,q) &=&\Pi_{D^*\bar{D} Z_c^{-}}(p^{\prime2},p^2,q^2)
\,i q_\alpha+\cdots\, ,
\end{eqnarray}

\begin{eqnarray}
	\Pi^{\chi_{c0}\pi }_{+}(p,q) &=&\Pi_{\chi_{c0}\pi Z_c^{+}}(p^{\prime2},p^2,q^2)
	+\cdots\, ,\nonumber\\[4pt]
	\Pi^{\eta_c a_0 }_{+}(p,q) &=&\Pi_{\eta_c a_0 Z_c^{+}}(p^{\prime2},p^2,q^2)
	+\cdots\, ,\nonumber\\[4pt]
	\Pi^{J/\psi\rho }_{+,\alpha\beta}(p,q) &=&\Pi_{J/\psi\rho Z_c^{+}}(p^{\prime2},p^2,q^2)
	\,\left(-i\varepsilon_{\alpha\beta\lambda\tau}p^\lambda q^\tau\right)+\cdots\, ,
\nonumber\\[4pt]
	\Pi^{D\bar{D}_0 }_{+}(p,q) &=&\Pi_{D\bar{D}_0 Z_c^{+}}(p^{\prime2},p^2,q^2)
		+\cdots\, ,
\nonumber\\[4pt]
	\Pi^{D^*\bar{D} }_{+,\alpha}(p,q) &=&\Pi_{D^*\bar{D} Z_c^{+}}(p^{\prime2},p^2,q^2)
\,i q_\alpha+\cdots\, ,
\nonumber\\[4pt]
	\Pi^{D^*\bar{D} }_{+,\alpha}(p,q) &=&\Pi_{D^*\bar{D} Z_c^{+}}(p^{\prime2},p^2,q^2)
\,i q_\alpha+\cdots\, ,
\nonumber\\[4pt]
	\Pi^{D^*\bar{D}^* }_{+,\alpha\beta}(p,q) &=&\Pi_{D^*\bar{D}^* Z_c^{+}}(p^{\prime2},p^2,q^2)
	\,\left(-i\varepsilon_{\alpha\beta\lambda\tau}p^\lambda q^\tau\right)+\cdots\, ,
\end{eqnarray}
where the decay constants and pole residues are defined by,
\begin{eqnarray}
\langle 0| J_\mu^{J/\psi}(0)|J/\psi(p)\rangle&=&f_{J/\psi} m_{J/\psi}\xi_\mu\, , \nonumber\\[4pt]
\langle 0| J_\alpha^{\chi_{c1}}(0)|\chi_{c1}(p)\rangle&=&f_{\chi_{c1}} m_{\chi_{c1}}\zeta_\alpha\, , \nonumber\\[4pt]
\langle 0| J^{\eta_c}(0)|\eta_c(p)\rangle&=&\frac{f_{\eta_c} m_{\eta_c}^2}{2m_c}\, ,\nonumber\\[4pt]
\langle 0| J^{\chi_{c0}}(0)|\chi_{c0}(p)\rangle&=&f_{\chi_{c0}} m_{\chi_{c0}}\, , \end{eqnarray}
\begin{eqnarray}
\langle0|J_{\mu}^{\rho}(0)|\rho(p)\rangle&=&f_{\rho} m_{\rho} \,\xi_\mu \,\, , \nonumber\\[4pt]
\langle 0| J_\alpha^{a_{1}}(0)|a_{1}(p)\rangle&=&f_{a_{1}} m_{a_{1}}\zeta_\alpha\, , \nonumber\\[4pt]
\langle0|J^{\pi}(0)|\pi(p)\rangle&=&\frac{f_{\pi} m_{\pi}^2}{m_u+m_d}  \,\, , \nonumber\\[4pt]
\langle 0| J^{a_0}(0)|a_0(q)\rangle&=&f_{a_0} m_{a_0}\, ,
\end{eqnarray}
\begin{eqnarray}
\langle 0| J_\alpha^{D^*}(0)|D^*(p)\rangle&=&f_{D^*} m_{D^*}\xi_\alpha\, , \nonumber\\[4pt]
\langle 0| J_\alpha^{D_1}(0)|D_1(p)\rangle&=&f_{D_1} m_{D_1}\zeta_\alpha\, ,\nonumber\\[4pt]
\langle 0| J^{D}(0)|D(p)\rangle&=&\frac{f_D m_D^2}{m_c}\, , \nonumber\\[4pt]
\langle 0| J^{D_0}(0)|D_0(p)\rangle&=&f_{D_0} m_{D_0}\, ,
\end{eqnarray}
and the hadronic coupling constants are defined by,
\begin{eqnarray}
\langle \chi_{c1}(p)\rho(q)|Z_c^{-}(p^\prime)\rangle&=&  \xi^*\cdot\zeta^* G_{\chi_{c1}\rho Z_c^{-}}\, , \nonumber\\[4pt]
\langle \eta_{c}(p)\rho(q)|Z_c^{-}(p^\prime)\rangle&=& i  \xi^* \cdot p \, G_{\eta_{c}\rho Z_c^{-}}\, ,\nonumber\\[4pt]
\langle J/\psi(p)a_1(q)|Z_c^{-}(p^\prime)\rangle&=&  \xi^* \cdot \zeta^* \, G_{J/\psi a_1 Z_c^{-}}\, ,\nonumber\\[4pt]
\langle J/\psi(p)\pi(q)|Z_c^{-}(p^\prime)\rangle&=& i  \xi^* \cdot q \,G_{J/\psi\pi Z_c^{-}}\, ,
\end{eqnarray}
\begin{eqnarray}
\langle D(p)\bar{D}_0(q)|Z_c^{-}(p^\prime)\rangle&=&  G_{D\bar{D}_0 Z_c^{-}}\, ,\nonumber\\[4pt]
\langle D^*(p)\bar{D}_1(q)|Z_c^{-}(p^\prime)\rangle&=&  \xi^* \cdot \zeta^* \, G_{D^*\bar{D}_1 Z_c^{-}}\, ,\nonumber\\[4pt]
\langle D^*(p)\bar{D}(q)|Z_c^{-}(p^\prime)\rangle&=& i  \xi^* \cdot q \,G_{D^*\bar{D} Z_c^{-}}\, ,
\end{eqnarray}
\begin{eqnarray}
	\langle \chi_{c0}(p)\pi(q)|Z_c^{+}(p^\prime)\rangle&=&  G_{\chi_{c0}\pi Z_c^{+}}\, ,\nonumber\\[4pt]
	\langle \eta_c(p)a_0(q)|Z_c^{+}(p^\prime)\rangle&=& i G_{\eta_c a_0 Z_c^{+}}\, ,\nonumber\\[4pt]
	\langle J/\psi(p)\rho(q)|Z_c^{+}(p^\prime)\rangle&=&  \varepsilon^{\lambda\tau\mu\nu}
	p_\lambda \xi^*_\tau q_\mu \xi^*_\nu \,G_{J/\psi\rho Z_c^{+}}\, ,
\end{eqnarray}
\begin{eqnarray}
	\langle D(p)\bar{D}_0(q)|Z_c^{+}(p^\prime)\rangle&=&  G_{D\bar{D}_0 Z_c^{+}}\, ,\nonumber\\[4pt]
	\langle D^*(p)\bar{D}_1(q)|Z_c^{+}(p^\prime)\rangle&=&  \xi^* \cdot \zeta^* \, G_{D^*\bar{D}_1 Z_c^{+}}\, ,\nonumber\\[4pt]
	\langle D^*(p)\bar{D}(q)|Z_c^{+}(p^\prime)\rangle&=& i  \xi^* \cdot q \,G_{D^*\bar{D} Z_c^{+}}\, ,\nonumber\\[4pt]
	\langle D^*(p)\bar{D}^*(q)|Z_c^{+}(p^\prime)\rangle&=&  \varepsilon^{\lambda\tau\mu\nu}
	p_\lambda \xi^*_\tau q_\mu \xi^*_\nu \,G_{D^*\bar{D}^* Z_c^{+}}\, ,
\end{eqnarray}
 the $\xi$ and $\zeta$ represent the polarization vectors of the vector and axialvector mesons, respectively, and the $\xi_\tau$ and $\xi_\nu$ represent the polarization vectors of the vector mesons $J/\psi$ ($D^*$) and $\rho$ ($\bar{D}^*$), respectively.
	
Through triple-dispersion relation, we can obtain the hadronic  spectral densities $\rho_H(s^\prime,s,u)$
\begin{eqnarray}\label{dispersion-3}
		\Pi_{H}(p^{\prime2},p^2,q^2)&=&\int_{4m_c^2}^\infty ds^{\prime} \int_{4m_c^2}^\infty ds \int_{0}^\infty du \frac{\rho_{H}(s^\prime,s,u)}{(s^\prime-p^{\prime2})(s-p^2)(u-q^2)}\, ,
\end{eqnarray}
where
\begin{eqnarray}
\rho_{H}(s^\prime,s,u)&=&{\lim_{\epsilon_3\to 0}}\,\,{\lim_{\epsilon_2\to 0}} \,\,{\lim_{\epsilon_1\to 0}}\,\,\frac{ {\rm Im}_{s^\prime}\, {\rm Im}_{s}\,{\rm Im}_{u}\,\Pi_{H}(s^\prime+i\epsilon_3,s+i\epsilon_2,u+i\epsilon_1) }{\pi^3} \, ,
\end{eqnarray}
we add the subscript $H$ to denote the hadron side.
		
On  the QCD side, we contract all the quark fields with the Wick's theorem,
 and obtain the results,
\begin{eqnarray}
\Pi(p,q)&\propto& \int d^4xd^4y e^{ip\cdot x}e^{iq \cdot y} \,{\rm Tr}\Big\{ \Gamma^a S_Q(0-x)\Gamma^b S_Q(x-0)\Gamma^c S^T_q(y-0)\Gamma^d S_q^T(0-y) \Big\} \, , \nonumber\\[3pt]
&{\rm or} & \int d^4xd^4y e^{ip\cdot x}e^{iq \cdot y}\, {\rm Tr}\Big\{\Gamma^a S^T_Q(x-0)\Gamma^b S_q^T(0-x) \Gamma^c S_Q(0-y)\Gamma^d S_q(y-0) \Big\} \, ,
\end{eqnarray}
where the $S_q$ and $S_Q$ are the (full) light and heavy quark propagators, respectively, the $\Gamma^{a/b/c/d}$ denote the Dirac $\gamma$-matrixes. There exist two loops, one connects  $0$ with $x$, the other connects  $0$ with $y$. If the vacuum condensates come from the quark-gluon operators of the same loop (different loops), the corresponding Feynman diagrams are disconnected (connected), which take place via the Okubo-Zweig-Iizuka super-allowed fall-apart mechanism without (with) exchanging gluons between the two color-neutral clusters. We take account of  both the connected and disconnected diagrams,
and carry out the operator product expansion up to the vacuum condensates of dimension 5, which contain the perturbative terms, quark condensates, gluon condensates  and quark-gluon mixed condensates. Generally speaking, in most cases, the main contributions come from the disconnected diagrams.  In some special cases, there only exist contributions come from the connected diagrams, which lead to small hadronic coupling constants, therefore small partial decay widths   \cite{WangZG-review,Wang:2017lot}.
Then we obtain the QCD spectral densities of the components $\Pi_{i}(p^{\prime2},p^2,q^2)$ through double-dispersion relation,
\begin{eqnarray}\label{dispersion-2}
\Pi_{QCD}(p^{\prime2},p^2,q^2)&=& \int_{\Delta_s^2}^\infty ds \int_{\Delta_u^2}^\infty du \frac{\rho_{QCD}(p^{\prime2},s,u)}{(s-p^2)(u-q^2)}\, ,
\end{eqnarray}
as
\begin{eqnarray}
{\rm lim}_{\epsilon_3 \to 0}{\rm Im}_{s^\prime}\,\Pi_{QCD}(s^\prime+i\epsilon_3,p^2,q^2)&=&0\, ,
\end{eqnarray}
with the thresholds $\Delta_s^2=4m_c^2$ or $m_c^2$, $\Delta_u^2=0$ or $m_c^2$.

On the hadron side, there is a triple dispersion relation, see Eq.\eqref{dispersion-3}, while on the QCD side, there is only a double dispersion relation, see Eq.\eqref{dispersion-2}. These relations cannot match with each other,  we firstly integrate over $ds^\prime$ on the hadron side,  then match the hadron side with the QCD side below the continuum thresholds $s_0$ and $u_0$ respectively to establish the quark-hadron duality rigorously \cite{Wang:2017lot,Wang:2019iaa},
\begin{eqnarray}\label{Duality}
\int_{\Delta_s^2}^{s_0}ds \int_{\Delta_u^2}^{u_0}du  \left[ \int_{4m_c^2}^{\infty}ds^\prime  \frac{\rho_H(s^\prime,s,u)}{(s^\prime-p^{\prime2})(s-p^2)(u-q^2)} \right] &=&\int_{\Delta_s^2}^{s_{0}}ds \int_{\Delta_u^2}^{u_0}du  \frac{\rho_{QCD}(s,u)}{(s-p^2)(u-q^2)}
		\, . \nonumber\\
\end{eqnarray}
For clearness,  we write down the hadron representation explicitly,
\begin{eqnarray}
	\Pi_{\chi_{c1}\rho Z_c^{-}}(p^{\prime2},p^2,q^2)&=&
	\frac{\lambda_{\chi_{c1}\rho Z_c^{-}}}{(m_{Z_{c}^{-}}^2-p^{\prime2})(m_{\chi_{c1}}^2-p^2)(m_\rho^2-q^2)}
	+\frac{C_{\chi_{c1}\rho Z_c^{-}}}{(m_{\chi_{c1}}^2-p^2)(m_\rho^2-q^2)}\nonumber\\
	&&+\cdots\, ,\nonumber\\[4pt]
	\Pi_{\eta_{c}\rho Z_c^{-}}(p^{\prime2},p^2,q^2)&=&
	\frac{\lambda_{\eta_{c}\rho Z_c^{-}}}{(m_{Z_{c}^{-}}^2-p^{\prime2})(m_{\eta_{c}}^2-p^2)(m_{\rho}^2-q^2)}
	+\frac{C_{\eta_{c}\rho Z_c^{-}}}{(m_{\eta_{c}}^2-p^2)(m_{\rho}^2-q^2)}\nonumber\\
	&&+\cdots\, ,\nonumber\\[4pt]
	\Pi_{J/\psi a_1 Z_c^{-}}(p^{\prime2},p^2,q^2)&=&
	\frac{\lambda_{J/\psi a_1 Z_c^{-}}}{(m_{Z_{c}^{-}}^2-p^{\prime2})(m_{J/\psi}^2-p^2)(m_{a_1}^2-q^2)}
	+\frac{C_{J/\psi a_1 Z_c^{-}}}{(m_{J/\psi}^2-p^2)(m_{a_1}^2-q^2)}\nonumber\\
	&&+\cdots\, , \nonumber\\[4pt]
	\Pi_{J/\psi\pi Z_c^{-}}(p^{\prime2},p^2,q^2)&=&
	\frac{\lambda_{J/\psi\pi Z_c^{-}}}{(m_{Z_{c}^{-}}^2-p^{\prime2})(m_{J/\psi}^2-p^2)(m_{\pi}^2-q^2)}
	+\frac{C_{J/\psi\pi Z_c^{-}}}{(m_{J/\psi}^2-p^2)(m_{\pi}^2-q^2)}\nonumber\\
	&&+\cdots\, ,
\end{eqnarray}

\begin{eqnarray}
	\Pi_{D\bar{D}_0 Z_c^{-}}(p^{\prime2},p^2,q^2)&=&
	\frac{\lambda_{D\bar{D}_0 Z_c^{-}}}{(m_{Z_{c}^{-}}^2-p^{\prime2})
		(m_{D}^2-p^2)(m_{\bar{D}_0}^2-q^2)}
	+\frac{C_{D\bar{D}_0 Z_c^{-}}}{(m_{D}^2-p^2)(m_{\bar{D}_0}^2-q^2)}\nonumber\\
	&&+\cdots\, ,\nonumber\\[4pt]
	\Pi_{D^*\bar{D}_1 Z_c^{-}}(p^{\prime2},p^2,q^2)&=&
	\frac{\lambda_{D^*\bar{D}_1 Z_c^{-}}}{(m_{Z_{c}^{-}}^2-p^{\prime2})
		(m_{D^*}^2-p^2)(m_{\bar{D}_1}^2-q^2)}
	+\frac{C_{D^*\bar{D}_1 Z_c^{-}}}{(m_{D^*}^2-p^2)(m_{\bar{D}_1}^2-q^2)}\nonumber\\
	&&+\cdots\, ,\nonumber\\[4pt]
	\Pi_{D^*\bar{D} Z_c^{-}}(p^{\prime2},p^2,q^2)&=&
	\frac{\lambda_{D^*\bar{D} Z_c^{-}}}{(m_{Z_{c}^{-}}^2-p^{\prime2})
		(m_{D^*}^2-p^2)(m_{\bar{D}}^2-q^2)}
	+\frac{C_{D^*\bar{D} Z_c^{-}}}{(m_{D^*}^2-p^2)(m_{\bar{D}}^2-q^2)}\nonumber\\
	&&+\cdots\, ,
\end{eqnarray}
\begin{eqnarray}
	\Pi_{\chi_{c0}\pi Z_c^{+}}(p^{\prime2},p^2,q^2)&=&
	\frac{\lambda_{\chi_{c0}\pi Z_c^{+}}}{(m_{Z_{c}^{+}}^2-p^{\prime2})
		(m_{\chi_{c0}}^2-p^2)(m_{\pi}^2-q^2)}
	+\frac{C_{\chi_{c0}\pi Z_c^{+}}}{(m_{\chi_{c0}}^2-p^2)(m_{\pi}^2-q^2)}\nonumber\\
	&&+\cdots\, ,\nonumber\\[4pt]
	\Pi_{\eta_c a_0 Z_c^{+}}(p^{\prime2},p^2,q^2)&=&
	\frac{\lambda_{\eta_c a_0 Z_c^{+}}}{(m_{Z_{c}^{+}}^2-p^{\prime2})
		(m_{\eta_c}^2-p^2)(m_{f_0}^2-q^2)}
	+\frac{C_{\eta_c a_0 Z_c^{+}}}{(m_{\eta_c}^2-p^2)(m_{f_0}^2-q^2)}\nonumber\\
	&&+\cdots\, ,\nonumber\\[4pt]
	\Pi_{J/\psi\rho Z_c^{+}}(p^{\prime2},p^2,q^2)&=&
	\frac{\lambda_{J/\psi\rho Z_c^{+}}}{(m_{Z_{c}^{+}}^2-p^{\prime2})
		(m_{J/\psi}^2-p^2)(m_{\rho}^2-q^2)}
	+\frac{C_{J/\psi\rho Z_c^{+}}}{(m_{J/\psi}^2-p^2)(m_{\rho}^2-q^2)}\nonumber\\
	&&+\cdots\, ,
\end{eqnarray}
\begin{eqnarray}
	\Pi_{D\bar{D}_0 Z_c^{+}}(p^{\prime2},p^2,q^2)&=&
	\frac{\lambda_{D\bar{D}_0 Z_c^{+}}}{(m_{Z_{c}^{+}}^2-p^{\prime2})
		(m_{D}^2-p^2)(m_{\bar{D}_0}^2-q^2)}
	+\frac{C_{D\bar{D}_0 Z_c^{+}}}{(m_{D}^2-p^2)(m_{\bar{D}_0}^2-q^2)}\nonumber\\
	&&+\cdots\, ,\nonumber\\[4pt]
	\Pi_{D^*\bar{D}_1 Z_c^{+}}(p^{\prime2},p^2,q^2)&=&
	\frac{\lambda_{D^*\bar{D}_1 Z_c^{+}}}{(m_{Z_{c}^{+}}^2-p^{\prime2})
		(m_{D^*}^2-p^2)(m_{\bar{D}_1}^2-q^2)}
	+\frac{C_{D^*\bar{D}_1 Z_c^{+}}}{(m_{D^*}^2-p^2)(m_{\bar{D}_1}^2-q^2)}\nonumber\\
	&&+\cdots\, ,
\nonumber\\[4pt]
	\Pi_{D^*\bar{D} Z_c^{+}}(p^{\prime2},p^2,q^2)&=&
	\frac{\lambda_{D^*\bar{D} Z_c^{+}}}{(m_{Z_{c}^{+}}^2-p^{\prime2})
		(m_{D^*}^2-p^2)(m_{\bar{D}}^2-q^2)}
	+\frac{C_{D^*\bar{D} Z_c^{+}}}{(m_{D^*}^2-p^2)(m_{\bar{D}}^2-q^2)}\nonumber\\
	&&+\cdots\, ,\nonumber\\[4pt]
	\Pi_{D^*\bar{D}^* Z_c^{+}}(p^{\prime2},p^2,q^2)&=&
	\frac{\lambda_{D^*\bar{D}^* Z_c^{+}}}{(m_{Z_{c}^{+}}^2-p^{\prime2})
		(m_{D^*}^2-p^2)(m_{\bar{D}^*}^2-q^2)}
	+\frac{C_{D^*\bar{D}^* Z_c^{+}}}{(m_{D^*}^2-p^2)(m_{\bar{D}^*}^2-q^2)}\nonumber\\
	&&+\cdots\, ,
\end{eqnarray}
where we introduce the parameters $C$ with different subscripts to stand for the contributions involving the higher resonances and continuum states in the $s^\prime$ channel \cite{Wang:2017lot,Wang:2019iaa,Wang:2023hsc,
Wang:2022fdu,Wang:2022ckc,Yang:2024guo,WangZG-WX-CPC-Pcdecay,
WangZG-WX-CPC-Pcdecay-Penta,WangZG-X3872-decay}.
				
We set $p^{\prime2}=\alpha p^2$ in the components $\Pi_H(p^{\prime 2},p^2,q^2)$, where the $\alpha$ is a finite quantity, and perform double Borel transformation  in regard  to the variables  $P^2=-p^2$ and $Q^2=-q^2$, respectively. If the final-state mesons are charmonium or bottomnium states, we can set $\alpha=1$. If the final-state mesons are open-charm or open-bottom mesons, we can set $\alpha=4$ \cite{Wang:2017lot}.

Then we set $T_1^2=T_2^2=T^2$  to obtain the QCD sum rules,
\begin{eqnarray}
&&\frac{\lambda_{\chi_{c1}\rho Z_c^-} G_{\chi_{c1}\rho Z_c^-}}{ m^2_{Z_c^-}-m^2_{\chi_{c1}}}\left[\exp\left(-\frac{m^2_{Z_c^-}}{T^2}\right) -\exp\left(-\frac{m^2_{\chi_{c1}}}{T^2}\right) \right] \exp\left(-\frac{m^2_\rho}{T^2}\right)\nonumber\\[3pt]
&&+C_{\chi_{c1}\rho Z_c^-} \exp\left(-\frac{m^2_{\chi_{c1}}}{T^2}-\frac{m^2_\rho}{T^2}\right)  \nonumber\\
&=&\frac{3}{32\sqrt{2}\pi^4} \int^{s_{\chi_{c1}}^0}_{4m^2_c} ds \int^{s_{\rho}^0}_{0} du \frac{ u\sqrt{s(s-4m^2_c)}^3}{s^2} \exp\left(-\frac{s+u}{T^2}\right) \nonumber\\[3pt]
&&+\frac{m^4_c}{6\sqrt{2}\pi^2}\langle\frac{\alpha_sGG}{\pi}\rangle \int^{s_{\chi_{c1}}^0}_{4m^2_c} ds \int^{s_\rho^0}_{0} du\frac{u}{\sqrt{s(s-4m^2_c)}^3} \exp\left(-\frac{s+u}{T^2}\right) \nonumber\\[3pt]
&&+\frac{1}{576\sqrt{2}\pi^2}\langle\frac{\alpha_sGG}{\pi}\rangle \int^{s_{\chi_{c1}}^0}_{4m^2_c} ds \int^{s_\rho^0}_{0} du\frac{s(u-4s)-8m^2_c(u-5s)}{s\sqrt{s (s-4m^2_c)}} \exp\left(-\frac{s+u}{T^2}\right) \nonumber\\[3pt]
&&+\frac{1}{96\sqrt{2}\pi^2}\langle\frac{\alpha_sGG}{\pi}\rangle \int^{s_{\chi_{c1}}^0}_{4m^2_c} ds\int^{s_\rho^0}_{0} du\frac{\sqrt{s(s-4m^2_c)}}{s} \exp\left(-\frac{s+u}{T^2}\right) \nonumber\\[3pt]
&&-\frac{m^4_c}{8\sqrt{2}\pi^2}\langle\frac{\alpha_sGG}{\pi}\rangle \int^{s_{\chi_{c1}}^0}_{4m^2_c} ds \int^{s_\rho^0}_{0} du\frac{su(s-2m^2_c)}{\sqrt{s(s-4 m^2_c)}^5} \exp\left(-\frac{s+u}{T^2}\right)  \ ,	 
\end{eqnarray}
where we introduce the notations,	
\begin{eqnarray}
	\lambda_{\chi_{c1}\rho Z_c^-}&=&{\lambda_{Z_c^-} f_{\chi_{c1}} m_{\chi_{c1}}f_{\rho} m_{\rho}}\, ,
\end{eqnarray}
and the other thirteen QCD sum rules are given explicitly in the Appendix. We take the unknown parameters $C$ as free parameters and adjust the suitable values to obtain flat Borel platforms for the hadronic coupling constants
 \cite{Wang:2017lot,Wang:2019iaa,Wang:2023hsc,Wang:2022fdu,Wang:2022ckc,
 WangZG-WX-CPC-Pcdecay,WangZG-WX-CPC-Pcdecay-Penta,WangZG-X3872-decay}. In calculations, we observe that there exist endpoint divergences at the thresholds  $s=4m_c^2$  due to powers of $s-4m_c^2$ in the denominators when the final-state mesons are charmonium states and $s=m_c^2$  due to powers of $s-m_c^2$ in the denominators when the final-state mesons are open-charm mesons. We add a shift term to remove the divergences via taking the replacements $s-4m_c^2\to s-4m_c^2+\Delta^2$ and $s-m_c^2\to s-m_c^2+\Delta^2$ with $\Delta^2=m_s^2$  \cite{Wang:2017lot,Yang:2024guo}.
	
\section{Numerical results and discussions}
On the QCD side, we take the standard vacuum  condensates $\langle \frac{\alpha_sGG}{\pi}\rangle=0.012\pm0.004\,\rm{GeV}^4$, $\langle \bar{q}q \rangle=-(0.24\pm 0.01\, \rm{GeV})^3$,
$\langle\bar{q}g_s\sigma G q \rangle=m_0^2\langle \bar{q}q \rangle$,
$m_0^2=(0.8 \pm 0.1)\,\rm{GeV}^2$ at the energy scale $\mu=1\, \rm{GeV}$  \cite{SVZ79-1,SVZ79-2,Reinders:1984sr,Colangelo:2000dp}
and take the $\overline{MS}$ mass $m_{c}(m_c)=(1.275\pm0.025)\,\rm{GeV}$  from the Particle Data Group \cite{ParticleDataGroup:2024cfk}. In addition,  we set $m_u=m_d=0$ and take account of the energy-scale dependence from re-normalization group equation,
\begin{eqnarray}
	\langle\bar{q}q \rangle(\mu)&=&\langle\bar{q}q \rangle({\rm 1GeV})\left[\frac{\alpha_{s}({\rm 1GeV})}{\alpha_{s}(\mu)}\right]^{\frac{12}{33-2n_f}}\, , \nonumber\\[3pt]
	\langle\bar{q}g_s \sigma Gq \rangle(\mu)&=&\langle\bar{q}g_s \sigma Gq \rangle({\rm 1GeV})\left[\frac{\alpha_{s}({\rm 1GeV})}{\alpha_{s}(\mu)}\right]^{\frac{2}{33-2n_f}}\, , \nonumber\\[3pt]
 m_c(\mu)&=&m_c(m_c)\left[\frac{\alpha_{s}(\mu)}{\alpha_{s}(m_c)}\right]^{\frac{12}{33-2n_f}} \, ,\nonumber\\[3pt]
	\alpha_s(\mu)&=&\frac{1}{b_0t}\left[1-\frac{b_1}{b_0^2}\frac{\log t}{t} +\frac{b_1^2(\log^2{t}-\log{t}-1)+b_0b_2}{b_0^4t^2}\right]\, ,
\end{eqnarray}
where $t=\log \frac{\mu^2}{\Lambda^2}$, $b_0=\frac{33-2n_f}{12\pi}$, $b_1=\frac{153-19n_f}{24\pi^2}$, $b_2=\frac{2857-\frac{5033}{9}n_f+\frac{325}{27}n_f^2}{128\pi^3}$,  $\Lambda=213\,\rm{MeV}$, $296\,\rm{MeV}$  and  $339\,\rm{MeV}$ for the flavors  $n_f=5$, $4$ and $3$, respectively \cite{ParticleDataGroup:2024cfk,Narison:1983kn}.
As we study the hidden-charm tetraquarks, we choose the flavor numbers $n_f=4$.

On the hadron side, we take the hadron masses $m_{\chi_{c1}}=3.51067\,\rm{GeV}$, $m_{\rho}=0.77526\,\rm{GeV}$, $m_{\eta_c}=2.9834\,\rm{GeV}$, $m_{J/\psi}=3.0969\,\rm{GeV}$, $m_{a_1}=1.23\,\rm{GeV}$, $m_{\pi}=0.13498\,\rm{GeV}$, $m_{\chi_{c0}}=3.41471\,\rm{GeV}$,  $m_{a_0}=0.980\,\rm{GeV}$, $m_{D}=1.86484\,\rm{GeV}$, $m_{D^*}= 2.00685\,\rm{GeV}$ from the Particle Data Group \cite{ParticleDataGroup:2024cfk}, $m_{D_0}=2.40\,\rm{GeV}$,  $m_{D_1}= 2.42\,\rm{GeV}$ \cite{Wang:2015mxa}, $m_{Z_{c}^{+}}=4.56\,\rm{GeV}$, and $m_{Z_{c}^{-}}=4.58 \,\rm{GeV}$ \cite{Wang:2021lkg} from the QCD sum rules.

And we take the decay constants or pole residues $f_{\chi_{c0}}=0.359\,\rm{GeV}$, $f_{\chi_{c1}}=0.338\,\rm{GeV}$ \cite{Novikov:1977dq}, $f_{\rho}=0.215\,\rm{GeV} $ \cite{Ball:2007rt}, $f_{\pi}=0.130\,\rm{GeV}$ \cite{Colangelo:2000dp}, $f_{\eta_c}=0.387\,\rm{GeV}$, $f_{J/\psi}=0.418\,\rm{GeV}$ \cite{Becirevic:2013bsa}, $f_{a_1}=0.238\,\rm{GeV}$ \cite{Wang:2008bw,Yang:2007zt},   $f_{a_0}=0.365\,\rm{GeV}$ \cite{Cheng:2022vbw,Wang:2015uha}, $f_{D}=0.208\,\rm{GeV}$, $f_{D_0}=0.373\,\rm{GeV}$, $f_{D^*}=0.263\,\rm{GeV}$, $f_{D_1}=0.332\,\rm{GeV}$ \cite{Wang:2015mxa}, $\lambda_{Z_c^{+}}=1.33 \times 10^{-1}\,\rm{GeV}^5$, $\lambda_{Z_c^{-}}=1.37 \times 10^{-1}\,\rm{GeV}^5$ \cite{Wang:2021lkg} from the QCD sum rules, and  $f_{\pi}m^2_{\pi}/(m_u+m_d)=-2\langle \bar{q}q\rangle/f_{\pi}$ from the Gell-Mann-Oakes-Renner relation.
Furthermore, we take the continuum threshold parameters   $s^0_{\chi_{c0}}=(3.9\,\rm{GeV})^2$, $s^0_{\chi_{c1}}=(4.0\,\rm{GeV})^2$ \cite{Novikov:1977dq}, $s^0_{\rho}=(1.2\,\rm{GeV})^2$ \cite{Ball:2007rt}, $s^0_{\pi}=(0.85\,\rm{GeV})^2$ \cite{Colangelo:2000dp}, $s^0_{\eta_c}=(3.5\,\rm{GeV})^2$, $s^0_{J/\psi}=(3.6\,\rm{GeV})^2$, $s^0_{a_1}=2.55\,\rm{GeV}^2$ \cite{Wang:2008bw},  $s^0_{a_0}=(1.2\,\rm{GeV})^2$, $s^0_{D}=6.2\,\rm{GeV}^2$, $s^0_{D_0}=8.3\,\rm{GeV}^2$, $s^0_{D^*}=6.4\,\rm{GeV}^2$, $s^0_{D_1}=8.6\,\rm{GeV}^2$ \cite{Wang:2015mxa} from the two-point QCD sum rules combined with the experimental data.

\begin{figure}
	\centering
	\includegraphics[width=0.9\columnwidth]{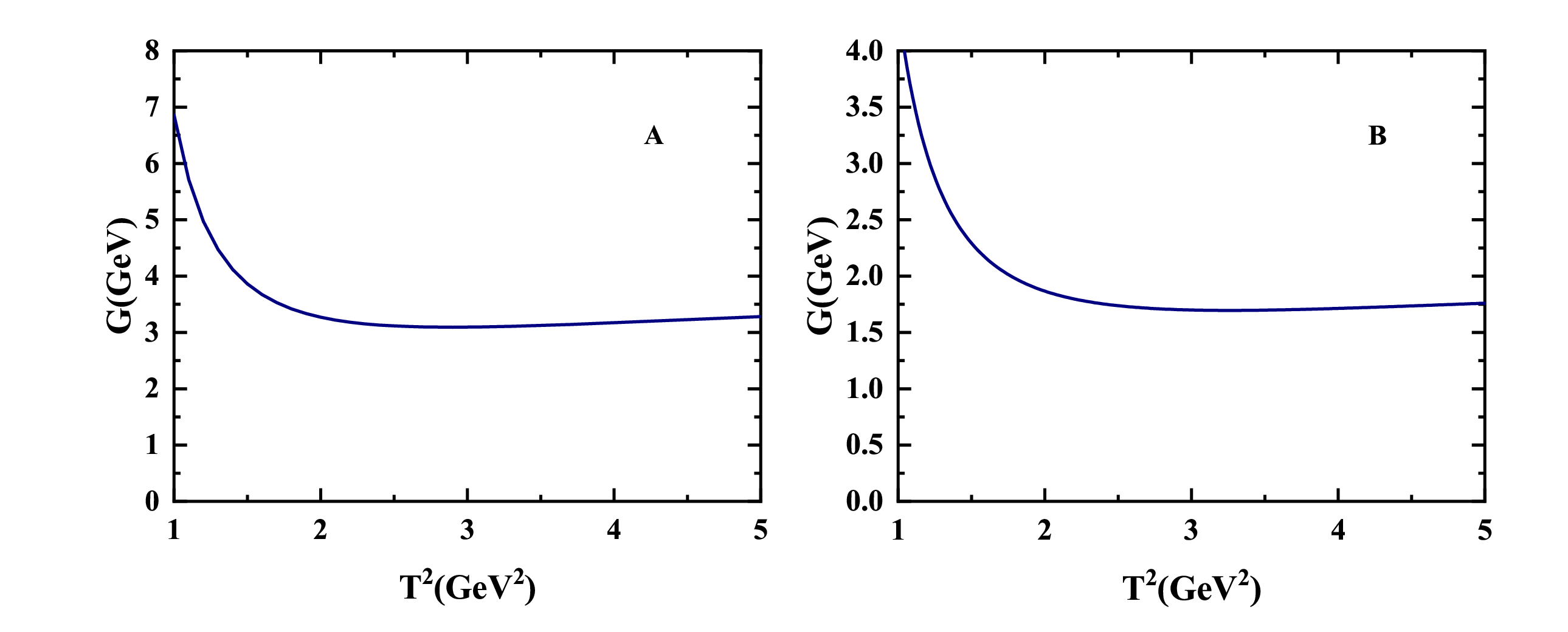}
	\caption{The hadronic coupling constants with variations of the Borel parameters, where the $A$ and $B$ denote the hadronic coupling constants $G_{\chi_{c1}\rho Z_c^{-}}$ and$G_{\chi_{c0}\pi Z_c^{+}}$, respectively.}\label{hadron-coupling-fig}
\end{figure}
The free parameters are fitted to obtain flat platforms, which are presented explicitly
in Table.\ref{tab:borelc}. Then we obtain uniform flat platforms  $T^2_{max}-T^2_{min}=1\,\rm{GeV}^2$ for all the channels, just like what have been done in our previous works \cite{WangZG-review,Wang:2017lot,Wang:2019iaa,Wang:2023hsc,Wang:2022fdu,Wang:2022ckc,Yang:2024guo}.

In Fig.\ref{hadron-coupling-fig}, the curves of the hadronic coupling constants  $G_{\chi_{c1} \rho Z_c^{-}}$ and $G_{\chi_{c0}\pi Z_c^{+}}$ are plotted with variations of the Borel parameters. There appear rather flat platforms  indeed, so it is reliable to extract the hadronic  coupling constants.

\begin{table}[htbp]
	\centering
	\small
	\setlength{\tabcolsep}{10pt}
	\renewcommand{\arraystretch}{1.3}
	\begin{tabular}{| c | c | c |}
		\hline
		Channels & $C$ & $T^2$ $\mathrm{GeV}^2$ \\
		\hline
		$\chi_{c1}\rho Z_c^{-}$      & $-0.00075~\mathrm{GeV}^6 \times T^2$ & 2.5--3.5 \\
		$\eta_c\rho Z_c^{-}$         & $+0.00049~\mathrm{GeV}^6 \times T^2$ & 1.5--2.5 \\
		$J/\psi a_1 Z_c^{-}$       & $-0.0039~\mathrm{GeV}^6 \times T^2$ & 3.0--4.0 \\
		$J/\psi\pi Z_c^{-}$          & $-0.00844~\mathrm{GeV}^6 \times T^2$ & 2.5--3.5 \\
		$D_0\bar{D}_0 Z_c^{-}$       & $-0.0155~\mathrm{GeV}^6 \times T^2$ & 1.5--2.5 \\
		$D^*\bar{D}_1 Z_c^{-}$       & $0$ & 4.0--5.0 \\
		$D^*\bar{D} Z_c^{-}$         & $-0.00165~\mathrm{GeV}^6 \times T^2$ & 2.0--3.0 \\
		\hline
		$\chi_{c0}\pi Z_c^{+}$       & $+0.0004~\mathrm{GeV}^6 \times T^2$ & 2.5--3.5 \\
		$\eta_c a_0 Z_c^{+}$         & $0$ & -- -- -- \\
		$J/\psi\rho Z_c^{+}$         & $0$ & 4.5--5.5 \\
		$D_0\bar{D}_0 Z_c^{+}$       & $-0.00325~\mathrm{GeV}^6 \times T^2$ & 2.0--3.0 \\
		$D^*\bar{D}_1 Z_c^{+}$       & $-0.0075~\mathrm{GeV}^6 \times T^2$ & 1.5--2.5 \\
		$D^*\bar{D} Z_c^{+}$         & $-0.00115~\mathrm{GeV}^6 \times T^2$ & 2.5--3.5 \\
		$D^*\bar{D}^* Z_c^{+}$         & $0$ & 2.0--3.0 \\
		\hline
	\end{tabular}
	\caption{The  free parameters $C$ and Borel platforms.}
	\label{tab:borelc}
\end{table}

Generally speaking, in the QCD sum rules, the uncertainties of the hadronic quantities, such as the masses, decay constants, coupling constants, originate from the quark masses, vacuum condensates, continuum threshold parameters and Borel parameters on the QCD side after all, we should take them into account in a consistent way.  In this work, we estimate the uncertainties in the same way as we usually do by setting $\delta \sqrt{s_0}=0$ for simplicity, as the central values of the hadronic quantities (especially the hadron masses) and continuum thresholds $\sqrt{s_0}$ have one to one correspondence \cite{WangZG-review,Wang:2023hsc,Wang:2022fdu,Wang:2022ckc,Yang:2024guo}.
We take the QCD sum
rules for the channel $Z_{c}^{-}\to \chi_{c1}\rho$ as an example, the uncertainties on the hadron side can be written as
\begin{eqnarray}
\lambda_{Z_c^-}f_{\chi_{c1}}f_{\rho}G_{\chi_{c1}\rho Z_c^-} &=& \bar{\lambda}_{Z_c^-}\bar{f}_{\chi_{c1}}\bar{f}_{\rho}\bar{G}_{\chi_{c1}\rho Z_c^-}
+\delta\,\lambda_{Z_c^-}f_{\chi_{c1}}f_{\rho}G_{\chi_{c1}\rho Z_c^-} \, , \nonumber\\[3pt]
C_{\chi_{c1}\rho Z_c^{-}}& =& \bar{C}_{\chi_{c1}\rho Z_c^-}+\delta C_{\chi_{c1} \rho Z_c^{-}} \, ,
 \end{eqnarray}
where
\begin{eqnarray}\label{Uncertainty-4}
	\delta\,\lambda_{Z_c^-}f_{\chi_{c1}}f_{\rho}G_{\chi_{c1}\rho Z_c^-} &=&\bar{\lambda}_{Z_c^-}\bar{f}_{\chi_{c1}}\bar{f}_{\rho} \bar{G}_{\chi_{c1}\rho Z_c^-}\left(\frac{\delta f_{\chi_{c1}}}{\bar{f}_{\chi_{c1}}}+\frac{\delta f_{\rho}}{\bar{f}_{\rho}} + \frac{\delta \lambda_{Z_c^-}}{\bar{\lambda}_{Z_c^-}} +\frac{\delta G_{\chi_{c1}\rho Z_c^-}}{\bar{G}_{\chi_{c1}\rho Z_c^-}}\right)\, ,
\end{eqnarray}
 the short overline denotes the central value.
It is obvious that the uncertainty $\delta\,\lambda_{Z}f_{\chi}f_{\rho}G$ originates  from uncertainties of all the input parameters on the QCD side as a collective effect via the formula,
\begin{eqnarray}
\delta f=\sqrt{\sum\limits_i\left[f(\bar x_i\pm \delta x_i)-f(\bar x_i)\right]^2}\, ,
\end{eqnarray}
where the $f(x_i)$ denotes the analytical expressions of the $\lambda_{Z}f_{\chi}f_{\rho}G$, and the input parameters $x_i=\bar{x}_i\pm \delta x_i$,
we could not distinguish the individual contributions $\delta f_{\chi}/\bar{f}_{\chi}$, $\delta f_{\rho}/\bar{f}_{\rho}$, $\delta \lambda_{Z}/\bar{\lambda}_{Z}$ and  $\delta G/\bar{G}$ unambiguously. The $\delta f_{\chi}/\bar{f}_{\chi}$, $\delta f_{\rho}/\bar{f}_{\rho}$ and $\delta \lambda_{Z}/\bar{\lambda}_{Z}$ differ significantly from the corresponding ones from the two-point QCD sum rules, as they are extracted from quite different Borel windows.
In calculations, we can set
\begin{eqnarray}
&&\delta C_{\chi_{c1}\rho Z_c^-}=0\, , \nonumber\\[3pt]
&&\frac{\delta f_{\chi_{c1}}}{\bar{f}_{\chi_{c1}}}=\frac{\delta f_{\rho}}{\bar{f}_{\rho}} = \frac{\delta \lambda_{Z_c^-}}{\bar{\lambda}_{Z_c^-}} =\frac{\delta G_{\chi_{c1}\rho Z_c^-}}{\bar{G}_{\chi_{c1}\rho Z_c^-}}\, ,
\end{eqnarray}
 approximately.
 In short, we take account of uncertainties of the quark masses, vacuum condensates and Borel parameters  consistently. As we adjust the unknown parameter $C$ to obtain flat platform, the uncertainty originates from the Borel parameter is less than $1\%$.

 If we insist on taking account of  the corresponding ones from the two-point QCD sum rules, we can add an additional uncertainty $\delta_A$,
 \begin{eqnarray}
 \delta_A&=&\sqrt{\left(\frac{\delta f_{\chi}}{\bar{f}_{\chi}}\right)^2+\left(\frac{\delta f_{\rho}}{\bar{f}_{\rho}}\right)^2 +\left(\frac{\delta \lambda_{Z}}{\bar{\lambda}_{Z}}\right)^2}\, ,
 \end{eqnarray}
 to the hadronic coupling constant $G$, i.e. $G\to G(1\pm\delta_A)$,  where the $\delta f_{\chi}$, $\delta f_{\rho}$ and $\delta \lambda_{Z}$ come from the two-point QCD sum rules considering  uncertainties of all the input parameters. Thus the partial decay width $\Gamma$ receives an additional uncertainty $2\delta_A$, as the $\Gamma\propto G^2$. We should bear in mind that the uncertainties are doubly counted.

 Finally, we obtain the values of the hadronic coupling constants,
 \begin{eqnarray} \label{HCC-values}
 G_{\chi_{c1}\rho Z_c^-} &=&3.10^{+0.38}_{-0.31}\,\rm{GeV}\, , \nonumber\\[4pt]
 G_{\eta_{c}\rho Z_c^-} &=&0.398^{+0.072}_{-0.062}\, , \nonumber\\[4pt]
 G_{J/\psi a_1 Z_c^-} &=&6.40^{+0.80}_{-0.67}\,\rm{GeV}\, , \nonumber\\[4pt]
 G_{J/\psi\pi Z_c^-}&=& 2.22^{+0.09}_{-0.10}\, ,\nonumber\\[4pt]
 G_{D\bar{D}_0 Z_c^-}&=& 2.98^{+0.25}_{-0.25}\,{\rm GeV},\nonumber\\[4pt]
 G_{D^*\bar{D}_1 Z_c^-}&=& 0.186^{+0.003}_{-0.004}\,{\rm GeV},\nonumber\\[4pt]
 G_{D^*\bar{D} Z_c^-}&=& 1.22^{+0.12}_{-0.12}\, ,
 \end{eqnarray}
 \begin{eqnarray}
 G_{\chi_{c0}\pi Z_c^+}&=& 1.70^{+0.30}_{-0.21}\,{\rm GeV}, \nonumber\\[4pt]
 G_{\eta_c a_0 Z_c^+}&=& 0\, , \nonumber\\[4pt]
 G_{J/\psi\rho Z_c^+}&=& 2.50^{+0.00}_{-0.00} \times 10^{-2}\,\rm{GeV}^{-1}, \nonumber\\[4pt]
 G_{D\bar{D}_0 Z_c^+}&=& 5.67^{+0.30}_{-0.30}\,{\rm GeV},\nonumber\\[4pt]
 G_{D^*\bar{D}_1 Z_c^+}&=& 3.26^{+0.33}_{-0.33}\,{\rm GeV},\nonumber\\[4pt]
 G_{D^*\bar{D} Z_c^+}&=& 0.34^{+0.13}_{-0.11}\, , \nonumber\\[4pt]
 G_{D^*\bar{D} Z_c^+}&=& 3.50^{+0.00}_{-0.00} \times 10^{-3}\,\rm{GeV}^{-1}\, .
 \end{eqnarray}
Then we obtain the partial decay widths directly,
\begin{eqnarray} \label{X0-decay-N}
\Gamma(Z_c^-\to \chi_{c1}\rho)&=& 49.42^{+0.74}_{-0.49} \,\rm{MeV}\, , \nonumber\\[4pt]
\Gamma(Z_c^-\to \eta_{c}\rho)&=& 1.067^{+0.035}_{-0.026} \,\rm{MeV}\, , \nonumber\\[4pt]
\Gamma(Z_c^-\to J/\psi a_1)&=& 195.97^{+3.06}_{-2.15} \,\rm{MeV}\, , \nonumber\\[4pt]
\Gamma(Z_c^-\to J/\psi\pi)&=& 38.76^{+0.06}_{-0.08} \,\rm{MeV}\, , \nonumber\\[4pt]	 
\Gamma(Z_c^-\to D\bar{D}_0)&=& 13.97^{+0.10}_{-0.10} \,\rm{MeV}\, , \nonumber\\[4pt]
\Gamma(Z_c^-\to D^*\bar{D}_1)&=& 0.13^{+0.00}_{-0.00} \,\rm{MeV}\, , \nonumber\\[4pt]	
\Gamma(Z_c^-\to D^*\bar{D})&=& 26.88^{+0.26}_{-0.26} \,\rm{MeV}\, ,
\end{eqnarray}
\begin{eqnarray} \label{X0-decay-P}
	\Gamma(Z_c^+\to \chi_{c0}\pi)&=& 5.55^{+0.17}_{-0.08} \,\rm{MeV}\, , \nonumber\\[4pt]
	\Gamma(Z_c^+\to \eta_c a_0)&=& 0.0 \,\rm{MeV}\, , \nonumber\\[4pt]	
	\Gamma(Z_c^+\to J/\psi\rho)&=& 5.50^{+0.00}_{-0.00} \times 10^{-2} \,\rm{MeV}\, , \nonumber\\[4pt]
	\Gamma(Z_c^+\to D\bar{D}_0)&=& 49.30^{+0.14}_{-0.14} \,\rm{MeV}\, , \nonumber\\[4pt]
	\Gamma(Z_c^+\to D^*\bar{D}_1)&=& 34.94^{+0.36}_{-0.36} \,\rm{MeV}\, , \nonumber\\[4pt]	
	\Gamma(Z_c^+\to D^*\bar{D})&=& 1.99^{+0.29}_{-0.21} \,\rm{MeV}\, , \nonumber\\[4pt]	
	\Gamma(Z_c^+\to D^*\bar{D}^*)&=& 8.47^{+0.00}_{-0.00} \times 10^{-5} \,\rm{MeV}\, .
\end{eqnarray}

Then we obtain the total decay  widths approximately,
\begin{eqnarray}
\Gamma(Z_c^-)&=&326.20^{+4.26}_{-3.11}\,\rm{MeV}\, ,  \nonumber\\[4pt]
\Gamma(Z_c^+)&=& 91.84^{+0.96}_{-0.76}\, {\rm MeV}\, .
\end{eqnarray}

We can easily determine the relative branching ratios of the pseudoscalar hidden-charm  tetraquark states from their partial decay widths,
\begin{align}
	&\Gamma \left( Z_c^- \to \chi_{c1} \rho : \eta_c \rho : J/\psi \pi : D\bar{D}_0 : D^* \bar{D}_1 : D^* \bar{D} : J/\psi a_1 \right)\nonumber\\[4pt]
	&= 0.252 : 0.000544 : 0.198 : 0.0713 : 0.000663 : 0.137 : 1.00 \, ,  \\[4pt]
	&\Gamma \left( Z_c^+ \to \chi_{c0} \pi : \eta_c a_0 : J/\psi \rho  : D^* \bar{D} : D^* \bar{D}_1: D^* \bar{D}^* : D\bar{D}_0 \right)\nonumber\\[4pt]
	&= 0.1125 : 0 : 0.001115 : 0.7088 : 0.04036 : 0 : 1.00\, .
\end{align}
Due to the particular quark structures, the dominant decay modes are $Z_c^- \to J/\psi a_1$ and $Z_c^+ \to D\bar{D}_0$, which could be observed experimentally in the future.

\section{Conclusion}
In this work, we study the hadronic coupling constants in the two-body strong decays of the  hidden-charm tetraquark states with the quantum numbers $J^{PC}=0^{-+}$and $0^{--}$  via the three-point correlation functions. We carry out the operator product expansion by considering  the quark condensates, gluon condensates and quark-gluon mixed condensates to obtain the QCD spectral representations, then match the QCD side with the hadron side  according to rigorous quark-hadron duality. We obtain the hadronic coupling constants and partial decay widths therefore total widths of the  hidden-charm tetraquark states with the $J^{PC}=0^{-+}$and $0^{--}$, respectively, which serve as a guide for the future experiments. Furthermore, we obtain the optimal channels $Z_c^- \to J/\psi a_1$ and $Z_c^+ \to D\bar{D}_0$ to search for the pseudoscalar hidden-charm tetraquark states experimentally in the future.
	
\section*{Appendix}
The analytical expressions of the other QCD sum rules,

\begin{eqnarray}
	&&\frac{\lambda_{\eta_c\rho Z_c^-} G_{\eta_c\rho Z_c^-}}{ m^2_{Z_c^-}-m^2_{\eta_c}}\left[\exp\left(-\frac{m^2_{Z_c^-}}{T^2}\right) -\exp\left(-\frac{m^2_{\eta_c}}{T^2}\right)\right] \exp\left(-\frac{m^2_\rho}{T^2}\right)\nonumber\\[3pt]
	&&+C_{\eta_c\rho Z_c^-} \exp\left(-\frac{m^2_{\eta_c}}{T^2}-\frac{m^2_\rho}{T^2}\right) \nonumber\\[3pt]
	&=&-\frac{3m_c}{16\sqrt{2}\pi^4} \int^{s_{\eta_c}^0}_{4m^2_c} ds \int^{s_\rho^0}_{0} du u \sqrt{1-\frac{4m^2_c}{s}} \exp\left(-\frac{s+u}{T^2}\right) \nonumber\\[3pt]
	&&+\frac{m_c}{24\sqrt{2}\pi^2}\langle\frac{\alpha_sGG}{\pi}\rangle \int^{s_{\eta_c}^0}_{4m^2_c} ds \int^{s_\rho^0}_{0} du\frac{1}{\sqrt{s(s-4m^2_c)}} \exp\left(-\frac{s+u}{T^2}\right) \nonumber\\[3pt]
	&&+\frac{3m_c}{64\sqrt{2}\pi^2}\langle\frac{\alpha_sGG}{\pi}\rangle \int^{s_{\eta_c}^0}_{4m^2_c} ds \int^{s_\rho^0}_{0} du\frac{u(5s^2
-14sm^2_c+4m^4_c)}{s\sqrt{s(s-4m^2_c)}^3} \exp\left(-\frac{s+u}{T^2}\right) \ ,	
\end{eqnarray}

\begin{eqnarray}
	&&\frac{\lambda_{J/\psi a_1 Z_c^-} G_{J/\psi a_1 Z_c^-}}{ m^2_{Z_c^-}-m^2_{J/\psi}}\left[\exp\left(-\frac{m^2_{Z_c^-}}{T^2}\right) -\exp\left(-\frac{m^2_{J/\psi}}{T^2}\right) \right] \exp\left(-\frac{m^2_{a_1}}{T^2}\right)\nonumber\\[3pt]
	&&+C_{J/\psi a_1 Z_c^-} \exp\left(-\frac{m^2_{J/\psi}}{T^2}-\frac{m^2_{a_1}}{T^2}\right) \nonumber\\[3pt]
	&=&\frac{3}{32\sqrt{2}\pi^4} \int^{s_{J/\psi}^0}_{4m^2_c} ds \int^{s_{a_1}^0}_{0} du u \sqrt{s(s-4m^2_c)} \exp\left(-\frac{s+u}{T^2}\right) \nonumber\\[3pt]
	&&+\frac{7m^4_c}{24\sqrt{2}\pi^2}\langle\frac{\alpha_sGG}{\pi}\rangle \int^{s_{J/\psi}^0}_{4m^2_c} ds \int^{s_{a_1}^0}_{0} du  \frac{u}{\sqrt{s(s-4 m^2_c)}^3} \exp\left(-\frac{s+u}{T^2}\right) \nonumber\\[3pt]
	&&-\frac{1}{576\sqrt{2}\pi^2}\langle\frac{\alpha_sGG}{\pi}\rangle \int^{s_{J/\psi}^0}_{4m^2_c}ds\int^{s_{a_1}^0}_{0} du\frac{2m^2_c(su+7u^2)+s(7su-2 u^2)}{su \sqrt{s(s-4m^2_c)}} \exp\left(-\frac{s+u}{T^2}\right) \nonumber\\[3pt]
	&&+\frac{1}{768\sqrt{2}\pi^2}\langle\frac{\alpha_sGG}{\pi}\rangle \int^{s_{J/\psi}^0}_{4m^2_c} ds \int^{s_{a_1}^0}_{0} du\frac{11s-18m^2_c}{\sqrt{s(s-4 m^2_c)}} \exp\left(-\frac{s+u}{T^2}\right)   \ ,	 
\end{eqnarray}

\begin{eqnarray}	
	&&\frac{\lambda_{J/\psi \pi Z_c^-} G_{J/\psi \pi Z_c^-}}{ m^2_{Z_c^-}-m^2_{J/\psi}}\left[\exp\left(-\frac{m^2_{Z_c^-}}{T^2}\right) -\exp\left(-\frac{m^2_{J/\psi}}{T^2}\right) \right] \exp\left(-\frac{m^2_\pi}{T^2}\right) \nonumber\\[3pt]
	&&+C_{J/\psi \pi Z_c^-} \exp\left(-\frac{m^2_{J/\psi}}{T^2}-\frac{m^2_\pi}{T^2}\right)  \nonumber\\[3pt]
	&=&\frac{1}{2\sqrt{2}\pi^2}\left(\langle\bar{q}q\rangle +\frac{\langle\bar{q}g_s\sigma Gq\rangle}{8T^2}\right)\int_{4m_c^2}^{s^0_{J/\psi}} ds \sqrt{s(s-4 m^2_c)}\exp\left(-\frac{s}{T^2}\right)\nonumber\\[3pt]
    &&-\frac{\langle\bar{q}g_s\sigma Gq\rangle}{48\sqrt{2}\pi^2T^2}\int_{4m_c^2}^{s^0_{J/\psi}}ds (s+2 m^2_c)\sqrt{1-\frac{4m^2_c}{s}} \exp\left(-\frac{s}{T^2}\right)\nonumber\\[3pt]
	&&-\frac{\langle\bar{q}g_s\sigma Gq\rangle}{384\sqrt{2}\pi^2}\int_{4m_c^2}^{s^0_{J/\psi}}ds\frac{37s+3m^2_c}{\sqrt{s(s-4 m^2_c)}}\exp\left(-\frac{s}{T^2}\right) \ ,	
\end{eqnarray}

\begin{eqnarray}	
	&&\frac{\lambda_{D\bar{D}_0 Z_c^-} G_{D\bar{D}_0 Z_c^-}}{ m^2_{Z_c^-}-m^2_D}\left[\exp\left(-\frac{m^2_{Z_c^-}}{T^2}\right) -\exp\left(-\frac{m^2_D}{T^2}\right) \right] \exp\left(-\frac{m^2_{\bar{D}_0}}{T^2}\right)\nonumber\\[3pt]
	&&+C_{D\bar{D}_0 Z_c^-} \exp\left(-\frac{m^2_D}{T^2}-\frac{m^2_{\bar{D}_0}}{T^2}\right)  \nonumber\\[3pt]
	&=&\frac{3}{16\sqrt{2}\pi^4} \int^{s_D^0}_{m^2_c} ds \int^{s_{\bar{D}_0}^0}_{m^2_c} du \frac{(s-m^2_c)^2(u-m^2_c)^2 }{su} \exp\left(-\frac{s+u}{T^2}\right) \nonumber\\[3pt]
    &&+\frac{m_c\langle\bar{q}q\rangle}{2\sqrt{2}\pi^2}\int_{m^2_c}^{s^0_{\bar{D}_0}} du	\frac{(u-m^2_c)^2}{u}\exp\left(-\frac{u+m^2_c}{T^2}\right)\nonumber\\[3pt]
	&&-\frac{m_c\langle\bar{q}q\rangle}{\sqrt{2}\pi^2}\int_{m^2_c}^{s^0_D} ds
	\frac{(s-m^2_c)^2}{s}\exp\left(-\frac{s+m^2_c}{T^2}\right)\nonumber\\[3pt]
	&&+\frac{m_c\langle\bar{q}g_s\sigma Gq\rangle}{4\sqrt{2}\pi^2T^2}\left(1-\frac{m_c^2}{2T^2}\right)\int_{m^2_c}^{s^0_{\bar{D}_0}}du\frac{(u-m^2_c)^2}{u} \exp\left(-\frac{u+m^2_c}{T^2}\right)\nonumber\\[3pt]
    &&-\frac{m^3_c\langle\bar{q}g_s\sigma Gq\rangle}{64\sqrt{2}\pi^2}\int_{m^2_c}^{s^0_{\bar{D}_0}}du\frac{u-3 m^2_c}{u^2}\exp\left(-\frac{u+m^2_c}{T^2}\right)\nonumber\\[3pt]
	&&-\frac{m_c\langle\bar{q}g_s\sigma Gq\rangle}{4\sqrt{2}\pi^2T^2}\left(1+\frac{m_c^2}{2T^2}\right)\int_{m^2_c}^{s^0_D}ds\frac{(s-m^2_c)^2 }{s}\exp\left(-\frac{s+m^2_c}{T^2}\right)  \ ,
\end{eqnarray}

\begin{eqnarray}
	&&\frac{\lambda_{D^*\bar{D}_1 Z_c^-} G_{D^*\bar{D}_1 Z_c^-}}{ m^2_{Z_c^-}-m^2_{D^*}}\left[\exp\left(-\frac{m^2_{Z_c^-}}{T^2}\right) -\exp\left(-\frac{m^2_{D^*}}{T^2}\right) \right] \exp\left(-\frac{m^2_{\bar{D}_1}}{T^2}\right)\nonumber\\[3pt]
	&&+C_{D^*\bar{D}_1 Z_c^-} \exp\left(-\frac{m^2_{D^*}}{T^2}-\frac{m^2_{\bar{D}_1}}{T^2}\right)  \nonumber\\[3pt]
	&=&\frac{m^4_c}{96\sqrt{2}\pi^2}\langle\frac{\alpha_sGG}{\pi}\rangle \int^{s_{D^*}^0}_{m^2_c} ds\int^{s_{\bar{D}_1}^0}_{m^2_c} du \frac{(s+u-m^2_c) (3s-u)}{s^2 u^2}\exp\left(-\frac{s+u}{T^2}\right) \nonumber\\[3pt]
	&&-\frac{m^3_c\langle\bar{q}g_s\sigma Gq\rangle}{32\sqrt{2}\pi^2T^4}\int^{s_{\bar{D}_1}^0}_{m^2_c} du\frac{(u-m^2_c)^2}{u} \exp\left(-\frac{u+m^2_c}{T^2}\right)\nonumber\\[3pt]
	&&-\frac{m^3_c\langle\bar{q}g_s\sigma Gq\rangle}{48\sqrt{2}\pi^2}\int^{s_{\bar{D}_1}^0}_{m^2_c} du\frac{2u-3m^2_c}{u^2 }\exp\left(-\frac{u+m^2_c}{T^2}\right)\nonumber\\[3pt]
	&&-\frac{m^3_c\langle\bar{q}g_s\sigma Gq\rangle}{48\sqrt{2}\pi^2}\int^{s_{D^*}^0}_{m^2_c} ds\frac{3s-2m^2_c}{s^2}\exp\left(-\frac{s+m^2_c}{T^2}\right)  \ ,	
\end{eqnarray}

\begin{eqnarray}
	&&\frac{\lambda_{D^*\bar{D} Z_c^-} G_{D^*\bar{D} Z_c^-}}{ m^2_{Z_c^-}-m^2_{D^*}}\left[\exp\left(-\frac{m^2_{Z_c^-}}{T^2}\right) -\exp\left(-\frac{m^2_{D^*}}{T^2}\right) \right] \exp\left(-\frac{m^2_{\bar{D}}}{T^2}\right) \nonumber\\[3pt]
	&&+C_{D^*\bar{D} Z_c^-} \exp\left(-\frac{m^2_{D^*}}{T^2}-\frac{m^2_{\bar{D}}}{T^2}\right)  \nonumber\\[3pt]
	&=&\frac{3m_c}{16\sqrt{2}\pi^4} \int^{s_{D^*}^0}_{m^2_c} ds \int^{s_{\bar{D}}^0}_{m^2_c} du \frac{(s-m^2_c)^2 (u-m^2_c)^2}{su^2} \exp\left(-\frac{s+u}{T^2}\right) \nonumber\\[3pt]	 &&+\frac{m^2_c\langle\bar{q}q\rangle}{2\sqrt{2}\pi^2}\int^{s_{\bar{D}}^0}_{m^2_c} du	\frac{(u-m^2_c)^2}{u^2}\exp\left(-\frac{u+m^2_c}{T^2}\right)\nonumber\\[3pt]
	&&-\frac{\langle\bar{q}q\rangle}{2\sqrt{2}\pi^2}\int_{4m_c^2}^{s^0_{D^*}} ds	 \frac{(s-m^2_c)^2}{s}\exp\left(-\frac{s+m^2_c}{T^2}\right)\nonumber\\[3pt]
	&&-\frac{m_c}{16\sqrt{2}\pi^2}\langle\frac{\alpha_sGG}{\pi}\rangle \int^{s_{D^*}^0}_{m^2_c} ds\int^{s_{\bar{D}}^0}_{m^2_c} du \frac{s-{m^2_c}}{u^2}\exp\left(-\frac{s+u}{T^2}\right) \nonumber\\[3pt]
	&&+\frac{m_c}{16\sqrt{2}\pi^2}\langle\frac{\alpha_sGG}{\pi}\rangle \int^{s_{D^*}^0}_{m^2_c} ds\int^{s_{\bar{D}}^0}_{m^2_c} du \frac{(s-m^2_c)(s-3m^2_c)}{su^2}\exp\left(-\frac{s+u}{T^2}\right) \nonumber\\[3pt]
	&&+\frac{\langle\bar{q}g_s\sigma Gq\rangle}{64\sqrt{2}\pi^2}\int^{s_{\bar{D}}^0}_{m^2_c} du\frac{(u-m^2_c)^2(u+m^2_c)}{u^3}
     \left[3+\frac{u+3 m^2_c}{T^2}+\frac{m^2_c(u-3 m^2_c)}{T^4}\right] \exp\left(-\frac{u+m^2_c}{T^2}\right)\nonumber\\[3pt]
	&&-\frac{\langle\bar{q}g_s\sigma Gq\rangle}{8\sqrt{2}\pi^2T^2}\left(2+\frac{m^2_c}{T^2}\right)
     \int^{s_{D^*}^0}_{m^2_c} ds\frac{(s-m^2_c)^2}{s}\exp\left(-\frac{s+m^2_c}{T^2}\right)\nonumber\\[3pt]
	&&+\frac{m^2_c\langle\bar{q}g_s\sigma Gq\rangle}{4\sqrt{2}\pi^2T^2}\int^{s_{\bar{D}}^0}_{m^2_c} du \frac{(u-m^2_c)^2}{ u^2}\exp\left(-\frac{u+m^2_c}{T^2}\right)  \ ,	
\end{eqnarray}

\begin{eqnarray}
	&&\frac{\lambda_{\chi_{c0}\pi Z_c^+} G_{\chi_{c0}\pi Z_c^+}}{ m^2_{Z_c^+}-m^2_{\chi_{c0}}}\left[\exp\left(-\frac{m^2_{Z_c^+}}{T^2}\right) -\exp\left(-\frac{m^2_{\chi_{c0}}}{T^2}\right) \right] \exp\left(-\frac{m^2_\pi}{T^2}\right)\nonumber\\[3pt]
	&&+C_{\chi_{c0}\pi Z_c^+} \exp\left(-\frac{m^2_{\chi_{c0}}}{T^2}-\frac{m^2_\pi}{T^2}\right)  \nonumber\\[3pt]
	&=&-\frac{3}{16\sqrt{2}\pi^4} \int^{s_{\chi_{c0}}^0}_{4m^2_c}ds\int^{s_\pi^0}_{0} du \frac{u \sqrt{s(s-4m^2_c)}^3}{s^2} \exp\left(-\frac{s+u}{T^2}\right) \nonumber\\[3pt]
	&&-\frac{m^2_c}{4\sqrt{2}\pi^2}\langle\frac{\alpha_sGG}{\pi}\rangle \int^{s_{\chi_{c0}}^0}_{4m^2_c} ds \int^{s_\pi^0}_{0} du\frac{u(s-3m_c^2)}{\sqrt{s(s-4 m^2_c)}^3} \exp\left(-\frac{s+u}{T^2}\right) \nonumber\\[3pt]
	&&+\frac{m^4_c}{4\sqrt{2}\pi^2}\langle\frac{\alpha_sGG}{\pi}\rangle \int^{s_{\chi_{c0}}^0}_{4m^2_c} ds \int^{s_\pi^0}_{0} du\frac{su (s-2m^2_c)}{\sqrt{s(s-4 m^2_c)}^5} \exp\left(-\frac{s+u}{T^2}\right) \nonumber\\[3pt]
	&&-\frac{m^4_c}{4\sqrt{2}\pi^2}\langle\frac{\alpha_sGG}{\pi}\rangle \int^{s_{\chi_{c0}}^0}_{4m^2_c} ds \int^{s_\pi^0}_{0} du\frac{u}{\sqrt{s(s-4m^2_c)}^3} \exp\left(-\frac{s+u}{T^2}\right) \ ,	
\end{eqnarray}

\begin{eqnarray}
	&&\frac{\lambda_{\eta_c a_0 Z_c^+} G_{\eta_c a_0 Z_c^+}}{ m^2_{Z_c^+}-m^2_{\eta_c}}\left[\exp\left(-\frac{m^2_{Z_c^+}}{T^2}\right) -\exp\left(-\frac{m^2_{\eta_c}}{T^2}\right) \right] \exp\left(-\frac{m^2_{a_0}}{T^2}\right)\nonumber\\[3pt]
	&&+C_{\eta_c a_0 Z_c^+} \exp\left(-\frac{m^2_{\eta_c}}{T^2}-\frac{m^2_{a_0}}{T^2}\right) \nonumber\\[3pt]
	&=& 0 \ ,	
\end{eqnarray}

\begin{eqnarray}
	&&\frac{\lambda_{J/\psi\rho Z_c^+} G_{J/\psi\rho Z_c^+}}{ m^2_{Z_c^+}-m^2_{J/\psi}}\left[\exp\left(-\frac{m^2_{Z_c^+}}{T^2}\right) -\exp\left(-\frac{m^2_{J/\psi}}{T^2}\right) \right] \exp\left(-\frac{m^2_\rho}{T^2}\right)\nonumber\\[3pt]
	&&+C_{J/\psi\rho Z_c^+} \exp\left(-\frac{m^2_{J/\psi}}{T^2}-\frac{m^2_\rho}{T^2}\right)  \nonumber\\[3pt]
	&=&\frac{31}{73728\sqrt{2}\pi^2}\langle\frac{\alpha_sGG}{\pi}\rangle \int^{s_{J/\psi}^0}_{4m^2_c} ds \int^{s_\rho^0}_{0} du \frac{\sqrt{s(s-4 m^2_c)}}{s^2} \exp\left(-\frac{s+u}{T^2}\right) \nonumber\\[3pt]
	&&+\frac{25}{73728\sqrt{2}\pi^2}\langle\frac{\alpha_sGG}{\pi}\rangle \int^{s_{J/\psi}^0}_{4m^2_c} ds \int^{s_\rho^0}_{0} du\frac{\sqrt{s(s-4m^2_c)}^3}{s^3u} \exp\left(-\frac{s+u}{T^2}\right) \nonumber\\[3pt]
	&&-\frac{17m^3_c\langle\bar{q}g_s\sigma Gq\rangle}{96\sqrt{2}\pi^2}\int^{s_{J/\psi}^0}_{4m^2_c} ds\frac{1}{s\sqrt{s(s-4 m^2_c)}}\exp\left(-\frac{s}{T^2}\right) \ ,	
\end{eqnarray}

\begin{eqnarray}	
	&&\frac{\lambda_{D\bar{D}_0 Z_c^+} G_{D\bar{D}_0 Z_c^+}}{ m^2_{Z_c^+}-m^2_D}\left[\exp\left(-\frac{m^2_{Z_c^+}}{T^2}\right) -\exp\left(-\frac{m^2_D}{T^2}\right) \right] \exp\left(-\frac{m^2_{\bar{D}_0}}{T^2}\right)\nonumber\\[3pt]
	&&+C_{D\bar{D}_0 Z_c^+} \exp\left(-\frac{m^2_D}{T^2}-\frac{m^2_{\bar{D}_0}}{T^2}\right)  \nonumber\\[3pt]
	&=&\frac{3m^2_c}{64\sqrt{2}\pi^4} \int^{s_D^0}_{m^2_c} ds \int^{s_{\bar{D}_0}^0}_{m^2_c} du\frac{(s-m^2_c)^2 (u-m^2_c)^2(3s-u) }{s^2u^2}\exp\left(-\frac{s+u}{T^2}\right) \nonumber\\[3pt]	 &&-\frac{m_c\langle\bar{q}q\rangle}{8\sqrt{2}\pi^2}\int^{s_{\bar{D}_0}^0}_{m^2_c} du \frac{(u-m^2_c)^2(u-3m^2_c)}{u^2}\exp\left(-\frac{u+m^2_c}{T^2}\right)\nonumber\\[3pt]
	&&-\frac{m_c\langle\bar{q}q\rangle}{8\sqrt{2}\pi^2} \int^{s_D^0}_{m^2_c} ds
	\frac{(s-m^2_c)^2(3 s-m^2_c)}{s^2}\exp\left(-\frac{s+m^2_c}{T^2}\right)\nonumber\\[3pt]
    &&-\frac{m^4_c}{64\sqrt{2}\pi^2}\langle\frac{\alpha_sGG}{\pi}\rangle \int^{s_D^0}_{m^2_c} ds \int^{s_{\bar{D}_0}^0}_{m^2_c} du\frac{(s+u-m^2_c)(3s-u)}{s^2u^2} \exp\left(-\frac{s+u}{T^2}\right) \nonumber\\[3pt]
	&&-\frac{m^2_c}{64\sqrt{2}\pi^2}\langle\frac{\alpha_sGG}{\pi}\rangle \int^{s_D^0}_{m^2_c} ds \int^{s_{\bar{D}_0}^0}_{m^2_c} du\frac{(s-m^2_c)(2u-m^2_c)(3s-u)}{s^2u^2} \exp\left(-\frac{s+u}{T^2}\right) \nonumber\\[3pt]
	&&-\frac{m_c\langle\bar{q}g_s\sigma Gq\rangle}{32\sqrt{2}\pi^2}\int_{m_c^2}^{s^0_{\bar{D}_0}}du
    \frac{(u-m^2_c)^2}{u^2}
    \left(4+\frac{4u-2m^2_c}{T^2}+\frac{3um^2_c}{T^4}\right)\exp\left(-\frac{u+m^2_c}{T^2}\right)\nonumber\\[3pt]
    &&-\frac{m_c\langle\bar{q}g_s\sigma Gq\rangle}{32\sqrt{2}\pi^2}\int_{m_c^2}^{s^0_D}ds\frac{(s-m^2_c)^2}{s^2}\left[3 +\frac{s+3 m^2_c}{T^2}+\frac{m^2_c(s-3 m^2_c)}{ T^4}\right]\exp\left(-\frac{s+m^2_c}{T^2}\right)\nonumber\\[3pt]
	&&+\frac{m^3_c\langle\bar{q}g_s\sigma Gq\rangle}{64\sqrt{2}\pi^2}\int_{m_c^2}^{s^0_{\bar{D}_0}}du\frac{u-3m^2_c}{u^2 }\exp\left(-\frac{u+m^2_c}{T^2}\right)\nonumber\\[3pt]
	&&+\frac{m^3_c\langle\bar{q}g_s\sigma Gq\rangle}{32\sqrt{2}\pi^2}\int_{m_c^2}^{s^0_D}ds
    \frac{3s-m^2_c}{s^2}\exp\left(-\frac{s+m^2_c}{T^2}\right)\nonumber\\[3pt]
	&&-\frac{m_c\langle\bar{q}g_s\sigma Gq\rangle}{32\sqrt{2}\pi^2}\int_{m_c^2}^{s^0_D}ds\frac{(s-m^2_c)^2}{s^2}\left(1+\frac{3 s-m^2_c}{T^2}\right)\exp\left(-\frac{s+m^2_c}{T^2}\right)  \ ,	
\end{eqnarray}

\begin{eqnarray}	
	&&\frac{\lambda_{D^*\bar{D}_1 Z_c^+} G_{D^*\bar{D}_1 Z_c^+}}{ m^2_{Z_c^+}-m^2_{D^*}}\left[\exp\left(-\frac{m^2_{Z_c^+}}{T^2}\right) -\exp\left(-\frac{m^2_{D^*}}{T^2}\right) \right] \exp\left(-\frac{m^2_{\bar{D}_1}}{T^2}\right)\nonumber\\[3pt]
	&&+C_{D^*\bar{D}_1 Z_c^+} \exp\left(-\frac{m^2_{D^*}}{T^2}-\frac{m^2_{\bar{D}_1}}{T^2}\right)  \nonumber\\[3pt]
	&=&\frac{3}{32\sqrt{2}\pi^4} \int^{s_{D^*}^0}_{m^2_c} ds \int^{s_{\bar{D}_1}^0}_{m^2_c} du \frac{(s-m^2_c)^2 (u-m^2_c)^2}{su}\exp\left(-\frac{s+u}{T^2}\right) \nonumber\\[3pt]
    &&+\frac{m_c\langle\bar{q}q\rangle}{4\sqrt{2}\pi^2}\int^{s_{\bar{D}_1}^0}_{m^2_c} du	\frac{(u-m^2_c)^2}{u}\exp\left(-\frac{u+m^2_c}{T^2}\right)\nonumber\\
	&&-\frac{m_c\langle\bar{q}q\rangle}{4\sqrt{2}\pi^2} \int^{s_{D^*}^0}_{m^2_c} ds	 \frac{(s-m^2_c)^2}{s}\exp\left(-\frac{s+m^2_c}{T^2}\right)\nonumber\\[3pt]
	&&+\frac{1}{2304\sqrt{2}\pi^2}\langle\frac{\alpha_sGG}{\pi}\rangle \int^{s_{D^*}^0}_{m^2_c} ds \int^{s_{\bar{D}_1}^0}_{m^2_c} du \frac{9s^2-19su+u^2}{s(u-m^2_c) }\exp\left(-\frac{s+u}{T^2}\right) \nonumber\\[3pt]
	&&+\frac{3m^{10}_c}{2304\sqrt{2}\pi^2}\langle\frac{\alpha_sGG}{\pi}\rangle \int^{s_{D^*}^0}_{m^2_c} ds \int^{s_{\bar{D}_1}^0}_{m^2_c} du \frac{36s^2-25su+4u^2}{s^3u^3 (u-m^2_c) }\exp\left(-\frac{s+u}{T^2}\right) \nonumber\\[3pt]
	&&-\frac{m^8_c}{2304\sqrt{2}\pi^2}\langle\frac{\alpha_sGG}{\pi}\rangle \int^{s_{D^*}^0}_{m^2_c} ds \int^{s_{\bar{D}_1}^0}_{m^2_c} du \frac{36s^3+65s^2u-41su^2+9 u^3}{s^3u^3(u-m^2_c) }\exp\left(-\frac{s+u}{T^2}\right) \nonumber\\[3pt]
	&&+\frac{m^6_c}{2304\sqrt{2}\pi^2}\langle\frac{\alpha_sGG}{\pi}\rangle \int^{s_{D^*}^0}_{m^2_c} ds \int^{s_{\bar{D}_1}^0}_{m^2_c} du \frac{36s^3-13s^2u+28su^2-9 u^3}{s^2u^3(u-m^2_c)}\exp\left(-\frac{s+u}{T^2}\right) \nonumber\\[3pt]
	&&-\frac{3m^4_c}{2304\sqrt{2}\pi^2}\langle\frac{\alpha_sGG}{\pi}\rangle \int^{s_{D^*}^0}_{m^2_c} ds \int^{s_{\bar{D}_1}^0}_{m^2_c} du \frac{9s^4-s^3u-16s^2u^2+7s u^3-u^4}{s^3u^2(u-m^2_c)}\exp\left(-\frac{s+u}{T^2}\right) \nonumber\\[3pt]
	&&-\frac{m^2_c}{2304\sqrt{2}\pi^2}\langle\frac{\alpha_sGG}{\pi}\rangle \int^{s_{D^*}^0}_{m^2_c} ds \int^{s_{\bar{D}_1}^0}_{m^2_c} du \frac{27s^2-16su+4u^2}{s^2 (u-m^2_c)}\exp\left(-\frac{s+u}{T^2}\right) \nonumber\\[3pt]
	&&+\frac{1}{96\sqrt{2}\pi^2}\langle\frac{\alpha_sGG}{\pi}\rangle \int^{s_{D^*}^0}_{m^2_c} ds\int^{s_{\bar{D}_1}^0}_{m^2_c} du \left(\frac{u-m^2_c}{u}-\frac{ s-m^2_c}{12s}\right)\exp\left(-\frac{s+u}{T^2}\right) \nonumber\\[3pt]
    &&+\frac{m^6_c}{1152\sqrt{2}\pi^2}\langle\frac{\alpha_sGG}{\pi}\rangle \int^{s_{D^*}^0}_{m^2_c} ds\int^{s_{\bar{D}_1}^0}_{m^2_c} du \frac{(s-m^2_c)(54s^2-39su+6 u^2)}{s^3u^3}\exp\left(-\frac{s+u}{T^2}\right) \nonumber\\[3pt]\notag
	&&+\frac{m^4_c}{1152\sqrt{2}\pi^2}\langle\frac{\alpha_sGG}{\pi}\rangle \int^{s_{D^*}^0}_{m^2_c}ds\int^{s_{\bar{D}_1}^0}_{m^2_c} du \frac{(s-m^2_c)(21s^2-16su+2 u^2)}{s^3u^2}\exp\left(-\frac{s+u}{T^2}\right)
\end{eqnarray}

\begin{eqnarray}	
	&&-\frac{m^2_c}{1152\sqrt{2}\pi^2}\langle\frac{\alpha_sGG}{\pi}\rangle \int^{s_{D^*}^0}_{m^2_c} ds\int^{s_{\bar{D}_1}^0}_{m^2_c} du \frac{ (s-m^2_c)(54s^3-34s^2u-7 su^2+2u^3)}{s^3u^2}\exp\left(-\frac{s+u}{T^2}\right) \nonumber\\[3pt]
	&&+\frac{1}{864\sqrt{2}\pi^2}\langle\frac{\alpha_sGG}{\pi}\rangle \int^{s_{D^*}^0}_{m^2_c} ds\int^{s_{\bar{D}_1}^0}_{m^2_c}du\frac{(2s^2-sm^2_c-m^4_c)(2u^2-um^2_c+3m^4_c)}{su^2 (u-m^2_c)}\exp\left(-\frac{s+u}{T^2}\right) \nonumber\\[3pt]
	&&-\frac{m^2_c}{864\sqrt{2}\pi^2}\langle\frac{\alpha_sGG}{\pi}\rangle \int^{s_{D^*}^0}_{m^2_c} ds\int^{s_{\bar{D}_1}^0}_{m^2_c}du\frac{(2s^2-sm^2_c-m^4_c)(2 u^2-um^2_c+3m^4_c)}{s^2u^2(u-m^2_c)}\exp\left(-\frac{s+u}{T^2}\right) \nonumber\\[3pt]
	&&-\frac{m^3_c\langle\bar{q}g_s\sigma Gq\rangle}{16\sqrt{2}\pi^2 T^4}\int^{s_{D^*}^0}_{m^2_c} ds\frac{(s-m^2_c)^2}{s}
    \exp\left(-\frac{s+m^2_c}{T^2}\right)\nonumber\\[3pt]
	&&-\frac{m_c\langle\bar{q}g_s\sigma Gq\rangle}{48\sqrt{2}\pi^2}\int^{s_{\bar{D}_1}^0}_{m^2_c} du\frac{2u+m^2_c}{u }\exp\left(-\frac{u+m^2_c}{T^2}\right)\nonumber\\[3pt]
	&&+\frac{m_c\langle\bar{q}g_s\sigma Gq\rangle}{48\sqrt{2}\pi^2}\int^{s_{D^*}^0}_{m^2_c} ds\frac{4s-m^2_c}{s}\exp\left(-\frac{s+m^2_c}{T^2}\right)  \ ,	
\end{eqnarray}

\begin{eqnarray}
	&&\frac{\lambda_{D^*\bar{D} Z_c^+} G_{D^*\bar{D} Z_c^+}}{ m^2_{Z_c^+}-m^2_{D^*}}\left[\exp\left(-\frac{m^2_{Z_c^+}}{T^2}\right) -\exp\left(-\frac{m^2_{D^*}}{T^2}\right) \right] \exp\left(-\frac{m^2_{\bar{D}}}{T^2}\right) \nonumber\\[3pt]
	&&+C_{D^*\bar{D} Z_c^+} \exp\left(-\frac{m^2_{D^*}}{T^2}-\frac{m^2_{\bar{D}}}{T^2}\right)  \nonumber\\[3pt]
	&=&-\frac{3m_c}{64\sqrt{2}\pi^4} \int^{s_{D^*}^0}_{m^2_c} ds \int^{s_{\bar{D}}^0}_{m^2_c} du \frac{(s-m^2_c)^2(u-m^2_c)^2(u+m^2_c)(3s-u)}{s^2u^3} \exp\left(-\frac{s+u}{T^2}\right) \nonumber\\[3pt]
	&&+\frac{\langle\bar{q}q\rangle}{2\sqrt{2}\pi^2}\int^{s_{\bar{D}}^0}_{m^2_c} du\frac{(u-m^2_c)^2(u+m^2_c)(u-3m^2_c)}{u^3}\exp\left(-\frac{u+m^2_c}{T^2}\right)\nonumber\\[3pt]
	&&-\frac{m_c}{384\sqrt{2}\pi^2}\langle\frac{\alpha_sGG}{\pi}\rangle \int^{s_{D^*}^0}_{m^2_c} ds\int^{s_{\bar{D}}^0}_{m^2_c}du\frac{(s+m^2_c)(u^2+3m^4_c)(3s-u)}{s^2 u^3}\exp\left(-\frac{s+u}{T^2}\right) \nonumber\\[3pt]
	&&-\frac{m_c}{384\sqrt{2}\pi^2}\langle\frac{\alpha_sGG}{\pi}\rangle \int^{s_{D^*}^0}_{m^2_c} ds\int^{s_{\bar{D}}^0}_{m^2_c} du \frac{(3s-5m^2_c)(u^2+2um^2_c-3m^4_c)(3s-u)}{s^2 u^3}\exp\left(-\frac{s+u}{T^2}\right) \nonumber\\[3pt]
	&&-\frac{m^5_c}{64\sqrt{2}\pi^2}\langle\frac{\alpha_sGG}{\pi}\rangle \int^{s_{D^*}^0}_{m^2_c} ds\int^{s_{\bar{D}}^0}_{m^2_c} du \frac{(s-m^2_c)(3s-u)}{s^2u^3 }\exp\left(-\frac{s+u}{T^2}\right) \nonumber\\[3pt]
	&&+\frac{m^2_c\langle\bar{q}g_s\sigma Gq\rangle}{64\sqrt{2}\pi^2T^4}\int^{s_{\bar{D}}^0}_{m^2_c} du\frac{(u-m^2_c)^2(u^2+6um^2_c-3m^4_c)}{u^3}
    \exp\left(-\frac{u+ m^2_c}{T^2}\right)\nonumber\\[3pt]
	&&+\frac{\langle\bar{q}g_s\sigma Gq\rangle}{64\sqrt{2}\pi^2}\int^{s_{\bar{D}}^0}_{m^2_c} du\frac{(u-m^2_c)^2 (u+m^2_c)}{u^3}\left(9+\frac{u-9 m^2_c}{T^2}\right)\exp\left(-\frac{u+m^2_c}{T^2}\right)\nonumber\\[3pt]
	&&-\frac{\langle\bar{q}g_s\sigma Gq\rangle}{96\sqrt{2}\pi^2}\int^{s_{\bar{D}}^0}_{m^2_c} du\frac{(u-3m^2_c)(2u^2+um^2_c+3m^4_c)}{u^3}\exp\left(-\frac{u+m^2_c}{T^2}\right)  \ ,	
\end{eqnarray}

\begin{eqnarray}
	&&\frac{\lambda_{D^*\bar{D}^* Z_c^+} G_{D^*\bar{D}^* Z_c^+}}{ m^2_{Z_c^+}-m^2_{D^*}}\left[\exp\left(-\frac{m^2_{Z_c^+}}{T^2}\right) -\exp\left(-\frac{m^2_{D^*}}{T^2}\right) \right] \exp\left(-\frac{m^2_{\bar{D}^*}}{T^2}\right) \nonumber\\[3pt]
	&&+C_{D^*\bar{D}^* Z_c^+} \exp\left(-\frac{m^2_{D^*}}{T^2}-\frac{m^2_{\bar{D}^*}}{T^2}\right)  \nonumber\\[3pt]
	&=&-\frac{17 }{73728\sqrt{2}\pi^2}\langle\frac{\alpha_sGG}{\pi}\rangle \int^{s_{D^*}^0}_{m^2_c} ds\int^{s_{\bar{D}^*}^0}_{m^2_c} du \frac{(2s-3 m^2_c)(u^2-m^4_c)  (3s-u)}{s^3u^2}\exp\left(-\frac{s+u}{T^2}\right) \nonumber\\[3pt]
	&&+\frac{1}{73728\sqrt{2}\pi^2}\langle\frac{\alpha_sGG}{\pi}\rangle \int^{s_{D^*}^0}_{m^2_c} ds \int^{s_{\bar{D}^*}^0}_{m^2_c} du \frac{(7s-28m^2_c)(3s-u)}{s^2 (u-m^2_c)}\exp\left(-\frac{s+u}{T^2}\right) \nonumber\\[3pt]
	&&+\frac{m^8_c}{73728\sqrt{2}\pi^2}\langle\frac{\alpha_sGG}{\pi}\rangle \int^{s_{D^*}^0}_{m^2_c} ds \int^{s_{\bar{D}^*}^0}_{m^2_c} du \frac{(576 m^2_c-555 u)(3 s-u) }{s^3u^3(u-m^2_c)}\exp\left(-\frac{s+u}{T^2}\right) \nonumber\\[3pt]
	&&+\frac{4m^6_c}{73728\sqrt{2}\pi^2}\langle\frac{\alpha_sGG}{\pi}\rangle \int^{s_{D^*}^0}_{m^2_c} ds \int^{s_{\bar{D}^*}^0}_{m^2_c} du \frac{(48s-7u)(3s-u)}{s^2u^3  (u-m^2_c)}\exp\left(-\frac{s+u}{T^2}\right) \nonumber\\[3pt]
	&&-\frac{m^4_c}{73728\sqrt{2}\pi^2}\langle\frac{\alpha_sGG}{\pi}\rangle \int^{s_{D^*}^0}_{m^2_c} ds \int^{s_{\bar{D}^*}^0}_{m^2_c} du \frac{(185s^2-21u^2)(3s-u) }{s^3u^2(u-m^2_c)}\exp\left(-\frac{s+u}{T^2}\right) \nonumber\\[3pt]
	&&-\frac{7 m^2_c}{36864\sqrt{2}\pi^2}\langle\frac{\alpha_sGG}{\pi}\rangle \int^{s_{D^*}^0}_{m^2_c} ds\int^{s_{\bar{D}^*}^0}_{m^2_c} du \frac{(s-m^2_c)(s-3 m^2_c)(3 s-u)}{s^3u^2}\exp\left(-\frac{s+u}{T^2}\right) \nonumber\\[3pt]
	&&+\frac{m^2_c}{288\sqrt{2}\pi^2}\langle\frac{\alpha_sGG}{\pi}\rangle \int^{s_{D^*}^0}_{m^2_c} ds \int^{s_{\bar{D}^*}^0}_{m^2_c} du \frac{(s-m^2_c) (u^2+2m^2_c-3m_c^4)(3s-u)}{s^3u^3}\exp\left(-\frac{s+u}{T^2}\right) , \nonumber\\[3pt]	
\end{eqnarray}
where we introduce the notations,
 \begin{eqnarray}
 	\lambda_{\eta_{c}\rho Z_c^-}&=&\frac{\lambda_{Z_c^-} f_{\eta_{c}} m^2_{\eta_{c}} f_{\rho} m_{\rho}}{2m_c}\, , \nonumber \\
 	\lambda_{J/\psi a_1 Z_c^-}&=&\lambda_{Z_c^-}f_{J/\psi} m_{J/\psi} f_{a_1} m_{a_1}\, , \nonumber \\
 	\lambda_{J/\psi\pi Z_c^-}&=&\frac{\lambda_{Z_c^-}f_{J/\psi} m_{J/\psi} f_{\pi} m_{\pi}^2}{m_u+m_d}\, , \nonumber \\
    \lambda_{D\bar{D}_0 Z_c^-}&=&\frac{\lambda_{Z_c^-}f_{{D}_0} m_{{D}_0} f_{D} m_{D}^2}{m_c}\, , \nonumber \\
    \lambda_{D^*\bar{D}_1 Z_c^-}&=&\lambda_{Z_c^-}f_{D} m_{D} f_{{D}_1} m_{{D}_1}\, , \nonumber \\
 	\lambda_{D^*\bar{D} Z_c^-}&=&\frac{\lambda_{Z_c^-}f_{D^*} m_{D^*} f_{D} m_{D}^2}{m_c}\, ,
 \end{eqnarray}

\begin{eqnarray}
	\lambda_{\chi_{c0}\pi Z_c^+}&=&\frac{\lambda_{Z_c^+} f_{\chi_{c0}} m_{\chi_{c0}} f_{\pi} m^2_{\pi}}{m_u+m_d}\, , \nonumber \\
	\lambda_{\eta_c a_0 Z_c^+}&=&\frac{\lambda_{Z_c^+} f_{\eta_c} m^2_{\eta_c} f_{a_0} m_{a_0}}{2m_c}\, , \nonumber \\
	\lambda_{J/\psi\rho Z_c^+}&=&\lambda_{Z_c^+}f_{J/\psi} m_{J/\psi} f_{\rho} m_{\rho}\, , \nonumber \\
	\lambda_{D\bar{D}_0 Z_c^+}&=&\frac{\lambda_{Z_c^+}f_{{D}_0} m_{{D}_0} f_{D} m_{D}^2}{m_c}\, , \nonumber \\
	\lambda_{D^*\bar{D}_1 Z_c^+}&=&\lambda_{Z_c^+}f_{D} m_{D} f_{{D}_1} m_{{D}_1}\, , \nonumber \\
	\lambda_{D^*\bar{D} Z_c^+}&=&\frac{\lambda_{Z_c^+}f_{D^*} m_{D^*} f_{D} m_{D}^2}{m_c}\, , \nonumber \\
	\lambda_{D^*\bar{D}^* Z_c^+}&=&\lambda_{Z_c^+}f_{D^*} m_{D^*}f_{D^*} m_{D^*} \, .
\end{eqnarray}

\section*{Acknowledgements}
This  work is supported by National Natural Science Foundation, Grant Number  12575083.

\end{document}